\documentclass[aps,showpacs]{revtex4-2}
\usepackage{amsfonts,amssymb,amsmath}
\usepackage{graphicx}
\usepackage{subfigure}
\usepackage[colorlinks=true, pdfstartview=FitV, linkcolor=blue, citecolor=red, urlcolor=magenta, breaklinks=true]{hyperref}
\begin{document}
\title{Oscillating soliton stars with network of domain walls}
\author{Stephen Owusu}
\email{stevenowusu15@yahoo.com.br}
\affiliation{Departamento de F\'{\i}sica, Universidade Federal de Campina Grande
Caixa Postal 10071, 58429-900 Campina Grande, Para\'{\i}ba, Brazil}
\affiliation{Instituto de F\'\i sica, Universidade Federal do Rio de Janeiro, Caixa Postal 21945, Rio de Janeiro, Rio de Janeiro, Brazil}

\author{Jos\'e A.V. Campos}
\email{joseandrecampos@gmail.com}
\affiliation{Departamento de F\'{\i}sica, Universidade Federal de Campina Grande
Caixa Postal 10071, 58429-900 Campina Grande, Para\'{\i}ba, Brazil}
\affiliation{Departamento de F\'\i sica, Universidade Federal da Para\'\i ba, Caixa Postal 5008, 58051-970 Jo\~ao Pessoa, Para\'\i ba, Brazil}

\author{Francisco A. Brito}
\email{fabrito@df.ufcg.edu.br}
\affiliation{Departamento de F\'{\i}sica, Universidade Federal de Campina Grande
Caixa Postal 10071, 58429-900 Campina Grande, Para\'{\i}ba, Brazil}
\affiliation{Departamento de F\'\i sica, Universidade Federal da Para\'\i ba, Caixa Postal 5008, 58051-970 Jo\~ao Pessoa, Para\'\i ba, Brazil}

\begin{abstract} 
In this work we study oscillating soliton stars (oscillatons) with network of domain walls. We consider a Lagrangian with three scalar fields coupled among themselves by a specific potential. We choose an appropriate potential to admit the formation of network of domain walls on the oscillatons. With small perturbations applied to this potential, we then compute the Einstein-Klein-Gordon (EKG) equations numerically and analyze the mass profile of this new object. From the results we discuss how the stability of the oscillatons is affected by the network.  At some conditions the network provides a `bouncing stability' to the oscillatons. 

\end{abstract}


\maketitle
\pretolerance10000

\section{INTRODUCTION}

A new class of astronomical objects was proposed by Friedberg, Lee, and Pang (1987) and called these objects soliton stars \cite{Lee,Lee and Pang,Friedberg}. These stars are objects of interest because they exhibit some remarkable properties. According to general relativity, if a star becomes massive enough, or captures enough matter greater than a critical value $M_c$, the star would undergo violent processes, either by expelling some of its mass and becoming a neutron star or a white dwarf  with $M<M_c$, or by collapsing into a black hole. Soliton star can exist with a very large stable cold mass configuration without becoming a black hole. Ordinary cold star like a white dwarf or a neutron star cannot have mass greater than five solar masses ($5M_\odot$ at zero angular momentum) \cite{C.W,R.M}. Soliton star is a type of stellar configuration that can have a very large mass ($>5M_\odot$), a very small volume and very high density. For example, a mini-soliton star could have a radius $6\times 10^{10} cm$, a mass of $10^{10} kg$ and density $10^{41}$ times that of a neutron star \cite{Friedberg2}.\

 In quantum field theory, when a discrete symmetry is spontaneously broken, for instance a $Z_2$ symmetry, a domain walls can form. For other $Z_N$ symmetries, with $N>2$ we have the possibility of having intersection of domain walls forming junctions, and then network of domain walls. In the context of soliton star there is the possibility for this network of domain walls to be entrapped on its surface, see \cite{Francisco}. For numerical realizations of such configurations in a spherical surface see \cite{sutcliffe}. Such network can appear for instance due to QCD phase transition which is governed by the $SU(3)$ group whose center is the discrete $Z_3$ symmetry. At the same scale, primordial black holes can also appear \cite{PBH,PBH-QCD,PBH-QCD2}. Thus, the network once formed may affect the stability of the soliton stars and can drive the rate of formation of such black holes.  \
 
 It has been long known that classical field theories admit non-topological soliton solution \cite{T.D}. These include Q-balls \cite{Coleman}, scalar soliton stars and boson stars \cite{E.S}. They are configurations made up of complex scalar fields, through nonlinear couplings of the scalar field to itself to other matter fields or to gravity. In the cases presented above, the conserved current is as a result of the global U(1) symmetry of the complex fields. But in the case of real scalar field, because of the absence of such symmetry there is no non-topological soliton solution. It can be shown that a massive real scalar field satisfying the Klein-Gordon equation can form a self-gravitating solitonic object when coupled to Einstein gravity. This new class of objects is not static, but rather periodic in time. We call such objects oscillating soliton stars \cite{W.M}. \
 
In this work we will investigate the formation of networks with the oscillating soliton star and study how the network affect the stability of the star by examining the mass profile. We execute this idea starting with appropriate model, described by three real scalar fields, introduced in line with \cite{D.F} --- see also \cite{stephen}.  An adjustment in the form of small perturbations applied to the potential will indicate where the network is formed with respect to the oscillating soliton star. This may give rise to three possibilities, that is having the network formed inside, or on the surface, or  outside the star. We analyze the stability of this new type of self-gravitating object and its significant role it can play in cosmology, as a dark matter candidate or as a significant source of gravitational wave. Since primordial black holes can also be candidates to dark matter \cite{PBH-DM} these new oscillating soliton stars (oscillatons) with network of domain walls can offer the possibility to uncover new scenarios. As source of gravitational waves some interesting new scenarios can also happen. For instance, rotating sources of two such objects can suffer transition to two black holes before they can be merged --- a signature that may appear in the detection of gravitational waves \cite{LIGO} and electromagnetic waves from the same event. As we shall show, the stability os such oscillatons is affected by slightly perturbing the network from their surface. Then approximating sources of such objects tidal forces can deform their surface in relation to the network destroying their stability driving them to collapse to black holes. \

This work will be organized as follows: In Sec.~\ref{sec1} we will introduce the basic aspects of the study of oscillating soliton star. Sec.~\ref{sec2} will contain the oscillating soliton star with network of domain walls. Finally Sec.~\ref{sec3} will hold comment, discussion and summary of the work.

\section{Oscillating Soliton Stars}
\label{sec1}
 
In this section let us quickly review the non self-interacting single scalar field model. 
The most complete action describing a self-gravitating real scalar field in curved space-time is
\begin{eqnarray}\label{action}
S= \int d^4x\sqrt{-g}\left[\frac{R}{2\kappa^2}+\mathcal{L}_m\right],
\end{eqnarray}  
where $\kappa^2=8\pi G$, $g=\det g_{\mu\nu}$, $R$ is the Ricci scalar, and $\mathcal{L}_m$ is the Lagrangian for matter field.  In this section, working with one scalar field we define the Lagrangian as
 \begin{eqnarray}
 \mathcal{L}_m=\frac{1}{2}[g^{\mu\nu}\partial_\mu\phi\partial_\nu\phi -V(\phi)],
 \end{eqnarray}
 where $\phi$ is the scalar field and $V(\phi)=m^2\phi^2$ is the potential when we consider free-field case with $m$ being the mass of the field. Variation of this action with respect to the scalar field gives the Klein-Gordon equation, as
\begin{eqnarray}\label{KG}
\square\phi=\frac{1}{2}\frac{\partial V}{\partial\phi}
\end{eqnarray}
being $\square\equiv\frac{1}{\sqrt{-g}}\partial_\mu(\sqrt{-g}g^{\mu\nu}\partial_\nu) $ the D'Alembertian in curved space. On the other hand, the variation of the action with respect to the metric $g^{\mu\nu}$ leads to the Einstein field equations given by 
\begin{eqnarray}
R_{\mu\nu}-\frac{1}{2}g_{\mu\nu}R= \kappa^2 T_{\mu\nu}\label{3.23}
\end{eqnarray}
where $ T_{\mu\nu}=\partial_{\mu}\phi\partial_{\nu}\phi -g_{\mu\nu}\mathcal{L}_m$ is the energy-momentum tensor.

Consider therefore a general time dependent, spherically symmetric geometry in a simple example of oscillating soliton stars, as a real massive Klein-Gordon scalar field, coupled to gravity as described in \cite{W.M,M. Al,Balakrishna}.  In the absence of angular momentum we consider the soliton solution to be spherically symmetric. The spherically symmetric line element is written in the form 
\begin{eqnarray}
ds^2=-N^2(t,r)dt^2 +g^2(t,r)dr^2 +r^2d\theta^2 +r^2\sin^2(\theta) d\varphi^2, \label{3.31}
\end{eqnarray}
with $N^2(t,r)$ being the lapse function and $ g^2(t,r)$ the radial metric function. Let us start by taking the case analyzed by \cite{W.M}, and write the coupled Einstein-Klein-Gordon (EKG) equations as:\\
\textit{the $t-t$ component}
\begin{eqnarray}
(g^2)'=-g^2\left(\frac{g^2-1}{r}\right)+4\pi Grg^2\left(\frac{\dot{\phi}g^2}{N^2}+\phi^{'2}+g^2m^2\phi^2\right),\label{3.43}
\end{eqnarray}
\textit{the $r-r$ component}
\begin{eqnarray}
(N^2)'=N^2(\frac{g^2-1}{r})+4\pi Gr(N^2\phi^{'2}-N^2g^2m^2\phi^2 +\dot{\phi}^2),\label{3.44}
\end{eqnarray}
\textit{the $t-r$ component}
\begin{eqnarray}
\dot{g}=4\pi Grg\dot{\phi}\phi'. \label{3.45}
\end{eqnarray}
We follow equation (\ref{KG}) to write the Klein-Gordon equation as
\begin{eqnarray}
\ddot{\phi}-\frac{\dot{N}\dot{\phi}}{N}=\frac{(N^2)'}{2g^2}\phi' +\frac{N^2}{g^2}\left[\phi'' -\frac{(g^2)'\phi'}{2g^2}-\frac{2g\dot{g}}{N^2}\right]+\frac{2N^2\phi'}{g^2r}-m^2N^2\phi ,\label{3.46}
\end{eqnarray}
where the over-dot represent $\frac{\partial}{\partial t}$ and the prime denote $\frac{\partial}{\partial r}$.
 For numerical convenience we define the dimensionless quantities $r\rightarrow r/m$, $t\rightarrow t/m$, $\Phi\rightarrow\frac{\phi}{\sqrt{\kappa^2}}$ where we note
that the bosonic mass is the natural scale for time and distance.
Also in order to deal with the non-linearity present in the EKG equations, it is convenient to introduce new variables  where $A(r,t)=g^2$ and $C=[g(r,t)/N(r,t)]^2$ \cite{M. Al,Balakrishna,L.A,L.A.2}. The EKG equations (\ref{3.43})-(\ref{3.46}) now becomes
\begin{eqnarray}
A'=-A\left(\frac{A-1}{r}\right)+\frac{Ar}{2}\left[C\dot{\Phi}^2+\Phi^{'2}+A\Phi^2\right],\label{3.47}
\end{eqnarray}
\begin{eqnarray}
C'=\frac{2C}{r}\left[1+A(\Phi^2 r^2-1)\right],\label{3.48}
\end{eqnarray}
\begin{eqnarray}
C\ddot{\Phi}+\frac{1}{2}\dot{C}\dot{\Phi}=\Phi'' +\Phi'\left(\frac{2}{r}-\frac{C'}{2C}\right)-A\Phi, \label{3.49}
\end{eqnarray}
\begin{eqnarray}
\dot{A}=2rA\dot{\Phi}\Phi'. \label{3.50}
\end{eqnarray}
These equations have no equilibrium solutions which are time independent (static metric component). All known static solutions to the system either have singularities or are topologically nontrivial \cite{T. Kodama} --- see also \cite{stability2}. The nature of equations (\ref{3.47})-(\ref{3.49}) suggests periodic expansion of the form 
 \begin{eqnarray}
 A(r,t)=\displaystyle\sum_{j= 0}^{\infty} A_{2j}(r)\cos(2j\omega t);
 \end{eqnarray}
\begin{eqnarray}
C(r,t)=\displaystyle\sum_{j= 0}^{\infty} C_{2j}(r)\cos(2j\omega t);
\end{eqnarray}
\begin{eqnarray}
\Phi(r,t)=\displaystyle\sum_{j= 1}^{\infty} \phi_{2j-1}(r)\cos[(2j-1)\omega t],
\end{eqnarray}
where $\omega$ is the fundamental frequency and will be absorbed in the redefinition of the time variable in the numerical calculations. The actual solution is an infinite Fourier expansion of the above form that is convergent \cite{W.M,M. Al,living}.
 
\subsection{Eigenvalue problem and the boundary conditions for equilibrium configurations}
\label{sec22}

We put these expansion into equations (\ref{3.47})-(\ref{3.49}), and set each Fourier coefficient to zero, to obtain a system of coupled nonlinear ordinary first-order differential equations for $A_{2j} (r)$ and $C_{2j} (r)$, and second order differential equations $\phi_{2j-1} (r)$. Eq.~(\ref{3.50}) is used as an algebraic equation to determine $A_0$.\

The boundary conditions are given by the following requirements:\\
i)	Asymptotic flatness:  this requires that as $ r\rightarrow\infty$, $A(r=\infty,t)=1$ and $\phi(r=\infty,t)=0$.  The coefficient $A_0(\infty)=1$, $A_j(\infty)=0$ for $j\neq 0$, $\phi_j(\infty)=0$ for all $j$, and $C_j(\infty)=0$ for $j\neq 0$. \\
ii)	Non-singularities: at $r=0$, the absence of a conical singularity implies $A_{2j} (r=0)=0$. The requirement that the metric coefficients be finite at $r=0$ implies $\phi' (r=0)=0$.\

The set of equations (\ref{3.47})-(\ref{3.49}) becomes an eigenvalue problem. Thus, it is necessary to determine the initial values $\phi_{2j-1}(0)$, $C_{2j}(0)$ and $\omega$, corresponding to a given central value $\phi_1(0)$. To proceed, we truncate the system of equations after a certain maximum $j=j_{max}=2$, numerically solve the eigenvalue problem by using the Runge-Kutta fourth-order method, and study the convergence of the series as function of $j=j_{max}$.

The soliton star total mass $M$. Asymptotically any soliton star (or boson)  metric resembles Schwarzschild metric, which allows us to associate the metric coefficient $g_{rr}=(1-2M/r)^{-1}$, where $M$ is the ADM mass defined for an asymptotically flat spacetime and $g_{rr}=g^2(r,t)=A(r,t)$. It can be calculated as
\begin{eqnarray}
 M=\lim_{r\to\infty}\frac{r}{2}\left[1-\frac{1}{A(r,t)}\right]\frac{M^2_{Planck}}{m}
\end{eqnarray}
where $r$ is the outermost point of numerical domain. For a recent discussion on this issue and boson stars, see for example \cite{ceres,Baibhav:2016fot}.\

\begin{figure}[h!]
\centering
		\includegraphics[width=10.0cm,height=6.6cm]{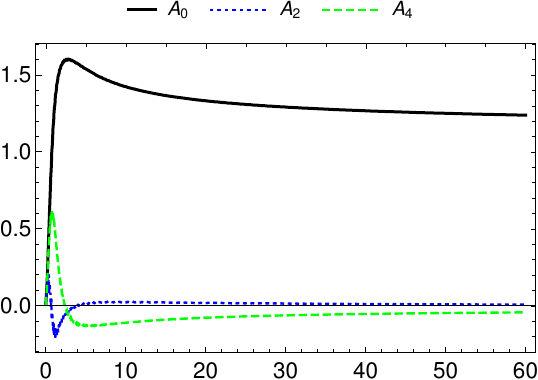}

\caption{A typical solution to the truncated eigenvalue equations ($j_{max}=2$) of the expansion of the metric $A(r,t)$. The solid, dotted and dashed lines show $A_0$, $A_2$ and $A_4$ respectively.}
\label{fig1}
\end{figure}

\begin{figure}[h!]
\centering
\includegraphics[width=10.0cm,height=6.6cm]{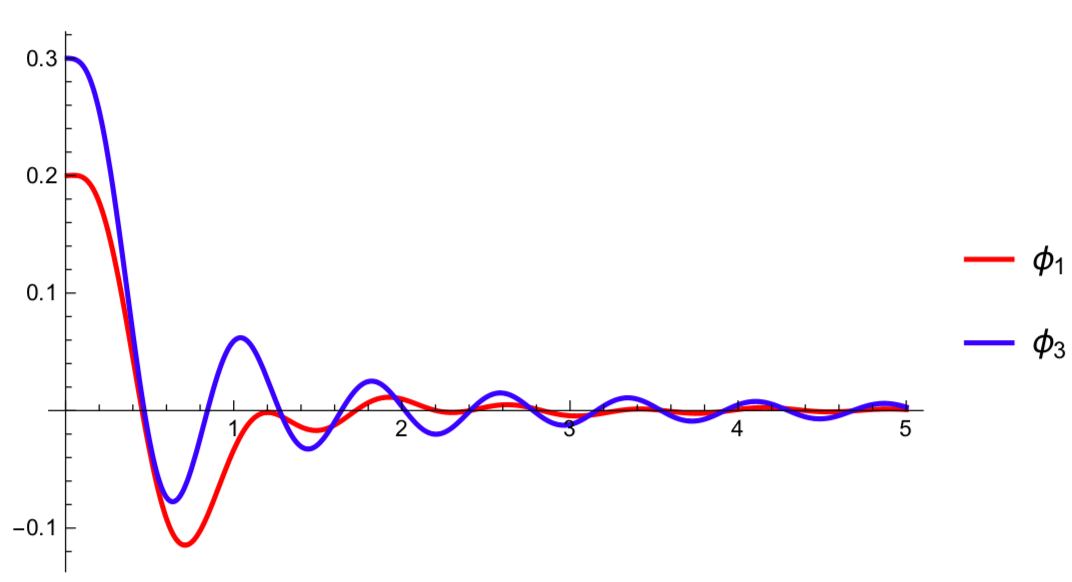}
\caption{The scalar field solution for the first two modes $\phi_1$ and $\phi_3$.}
\label{fig1-1}
\end{figure}

A typical oscillaton solution is shown in Fig.~\ref{fig1} and the scalar field modes are shown in Fig.~\ref{fig1-1}.  The radial metric function $A(r,t)$, for the case of $\phi_1(0)=0.2$ is plotted for individual  $A_{2j}$ for the first few values of $j$. 
Though we are solving non-linear equations, the Fourier series converges rapidly. \

\begin{figure}[h!]
\begin{center}
\includegraphics[scale=1]{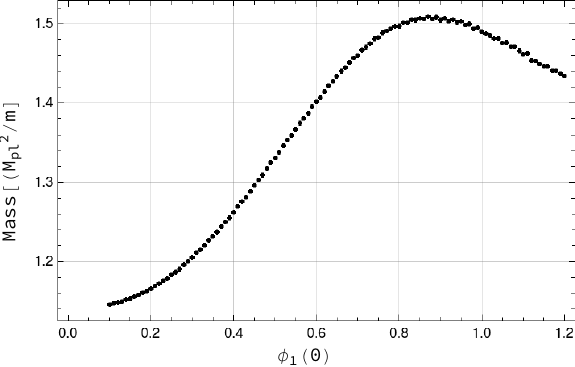}
\end{center}
\caption{The total mass $M$ of the oscillating soliton star (in units of $M^2_{Planck}/m$) is plotted as a function of central density $\phi(0)$ for stars in their first excited state $\phi_1(0)$. The circles represent actual configurations resulting from solutions to the eigenvalue equations.}
\label{fig2}
\end{figure}

In Fig.~\ref{fig2} the mass $M$ of the star is plotted as a function of the central field $\phi_1(0)$. This mass curve is similar to those of white dwarfs, neutron stars, and boson stars, with a maximum mass given by $M_c\approx {1.51}M^2_{Planck}/m$ at $\phi_{1c}(0)\approx {0.9}$. The branch to the left of the maximum is the stable branch traditionally called the S-branch. Stability here means that the stars on this branch move to new lower mass configurations on the same branch under small perturbations. To the right is the unstable branch called U-branch. U-branch are stars inherently unstable to small perturbations and collapse to black hole. In this regime, even small increase in mass will induce the collapse of the star into a black hole or lose of mass through scalar radiation  will cause it to migrate to the S-branch.

\section{Oscillating Soliton Stars with Network of Domain Walls}
\label{sec2}

Let us explore the idea of having a network of domain walls living on an oscillating soliton star. In this present work we offer a model that contains combined ideas of the basic mechanism seen in section \ref{sec1} and the ideas treated in Refs.~\cite{Francisco, D.F, stephen}. The model will ultimately lead to the scenario of oscillating soliton stars hosting  a network of domain walls (or defects). This will be done by employing a Lagrangian which is made up of three scalar fields that is coupled among themselves by a potential. In line with \cite{Francisco}, we will consider the Lagrangian to be
\begin{eqnarray}
\mathcal{L}_m=\frac{1}{2}\partial_\mu\phi\partial^\mu\phi +\frac{1}{2}\partial_\mu\chi\partial^\mu\chi +\frac{1}{2}\partial_\mu\sigma\partial^\mu\sigma -V(\phi ,\chi ,\sigma).\label{4.1}
\end{eqnarray}
This model is made up of three real scalar fields that is  coupled among themselves by the potential $V(\phi ,\chi ,\sigma)$. Putting this potential into the action given by equation (\ref{action}) and varying it with respect to the scalar fields $(\sigma, \phi, \chi)$ lead to the following Klein-Gordon equations
\begin{eqnarray}
  \frac{1}{\sqrt{-g}}\partial_\mu(\sqrt{-g}g^{\mu\nu}\partial_\nu\phi)-\frac{\partial V(\phi ,\chi ,\sigma)}{\partial\phi}=0,\label{kg46}
  \end{eqnarray} 
   \begin{eqnarray}
  \frac{1}{\sqrt{-g}}\partial_\mu(\sqrt{-g}g^{\mu\nu}\partial_\nu\chi)-\frac{\partial V(\phi ,\chi ,\sigma)}{\partial\chi}=0,\label{kg47}
  \end{eqnarray} 
   \begin{eqnarray}
  \frac{1}{\sqrt{-g}}\partial_\mu(\sqrt{-g}g^{\mu\nu}\partial_\nu\sigma)-\frac{\partial V(\phi ,\chi ,\sigma)}{\partial\sigma}=0.\label{kg48}
\end{eqnarray}
    
\subsection{The soliton star as a  bubble wall}
  Generally we are interested in the model that should be able to describe a spherical soliton star via the scalar field $\sigma$ (host field), by breaking its $Z_2$ symmetry under the shift $\sigma\rightarrow\sigma -\frac{1}{2}\sigma_0$, to entrap the other two fields ($\phi ,\chi$) with a $Z_3$ symmetry on its surface \cite{Francisco}. We obtain this by considering the potential of the form  
  \begin{eqnarray}
  V=\frac{1}{2}\mu^2\sigma^2(\sigma -\sigma_0)^2-{ \mu^2\varepsilon\sigma_0\sigma^3}+\lambda^2(\phi^2 +\chi^2)^2-\lambda^2\phi(\phi^2 -3\chi^2)+[\lambda\mu(\sigma -\frac{1}{2}\sigma_0)^2 -\frac{9}{4}\lambda^2](\phi^2 +\chi^2),\label{22}
  \end{eqnarray}
where $\sigma=0$ and $\sigma=\sigma_0$ are the false and true vacua corresponding to the standard soliton star. The $\varepsilon$-term is responsible to yield non-degenerate vacua in the potential which supports the existence of a spherical domain wall whose surface tension is the same as of a bubble wall \cite{Kolb}. At the thin wall limit, i.e. $\varepsilon\ll1$,  the radius $R$ of the bubble is very large and the bubble wall approaches an infinite planar domain wall.

\begin{figure}[h!]
\centering
		\includegraphics[width=12.0cm,height=8.8cm]{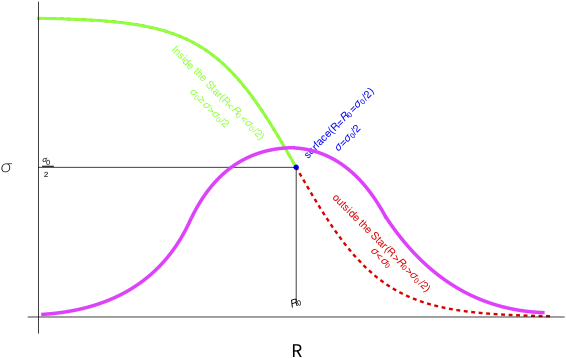}
\caption{The schematic behavior of the $\sigma$ field that connects true (inside) and false (outside) vacua. The ``bell-shaped'' curve stands for the fields $\phi, \chi$ responsible for the network on the soliton star. These fields go to zero at the vacua but develops a maximum at the surface of the spherical domain wall governed by the scalar field $\sigma$.}
\label{fig1-0}
\end{figure}

We use the equations of motion to see that in the true and false vacua $\sigma\approx\sigma_0$ and $\sigma=0$, respectively, the fields ($\phi, \chi$) turn out to be zero,  see Fig.~\ref{fig1-0}. For scalar fields $\phi$, $\chi=0$  the theory (\ref{22}) allows the field $\sigma$ to form a soliton solution. We note that at Minkowski space (or at very weak gravitational field) such a solution can be found by using the following first order differential equation
\begin{eqnarray}
\frac{d\sigma}{dR}=\mu\sigma(\sigma-\sigma_0)=W_\sigma, \label{4.9}
\end{eqnarray}
where $W_\sigma=\frac{dW}{d\sigma}$. The function $W=\mu(\sigma^3/3-\sigma^2\sigma_0/2)$ is well-known as the `superpotential' that defines the potential $V(\sigma ,0,0)=(1/2)W_\sigma^2$ \cite{D.To}. Integrating equation (\ref{4.9}) gives the solution
\begin{eqnarray}
\sigma=\frac{\sigma_0}{2}\left[1-\tanh\frac{\mu\sigma_0(R-R_0)}{2}\right].\label{4.11}
\end{eqnarray}
In the presence of gravitational field the numerical solution, say for the first mode, as presented in Fig.~\ref{fig1-1} approaches the kink profile \eqref{4.11} at the `thick' wall limit  \cite{Kolb} --- see also the schematic behavior of this solution in Fig.~\ref{fig1-0}. This solution indicates that at the surface of the star ($R\simeq R_0$) the sigma field goes to $(1/2)\sigma_0$. This represents approximately \cite{Kolb} a spherical wall (the soliton star surface) with surface tension \cite{Francisco}
\begin{eqnarray}
t_h\simeq |W(\sigma_0)-W(0)|=\frac{1}{6}\mu\sigma_0^3. \label{4.12}
\end{eqnarray}

When $\sigma\simeq (1/2)\sigma_0$ the other two scalar fields ($\phi ,\chi$) develops vacua, see Fig.~\ref{fig1-0}, that respects the $Z_3$ symmetry, and describe three-junctions of domain walls which allow the formation of a network on the surface \cite{D.F,Francisco}.  The solution also indicates that outside ($R>R_0$) and inside ($R<R_0$) the star the sigma field goes to $< (1/2)\sigma_0$ and $>(1/2)\sigma_0$ respectively.

\subsection{Einstein-Klein-Gordon equations for three scalar fields}
In this section, we will calculate the EKG equations for three scalar fields using the potential defined by equation (\ref{22}) in curved space. The soliton star solutions now can be found only numerically, as we shall see shortly.


The Einstein's equation for three scalar fields coupled with each other aided by a potential are obtained from the variation of the action given with respect to the metric tensor $g^{\mu\nu}$ that leads to
\begin{eqnarray}
 R_{\mu\nu}-\frac{1}{2}g_{\mu\nu}R=\kappa^2 T_{(\phi ,\chi ,\sigma)\mu\nu},\label{4.19}
 \end{eqnarray}
 where $T_{(\phi ,\chi ,\sigma)\mu\nu}$ is the energy-momentum tensor for the scalar fields. 
 
To proceed we will consider the following conditions. The constants are assumed $\lambda=\mu=1$ on the GeV scale \cite{Friedberg,Friedberg2} and the sigma  field is considered to be periodic in time, whereas the other two fields ($\phi$ and $\chi$) are static fields because they describe a co-moving network living around the oscillating boson star surface. The potential then becomes
\begin{eqnarray}
V=\frac{1}{2}\sigma^2(\sigma-\sigma_0)^2 +(\phi^2 +\chi^2)^2 -\phi (\phi^2 -3\chi^2)+\left[(\sigma-\frac{1}{2}\sigma_0)^2-\frac{9}{4}\right](\phi^2 +\chi^2).\label{4.15}
\end{eqnarray} 
  The general energy-momentum tensor associated with it is 
 \begin{equation}
 \begin{split}
  T_{(\phi ,\chi ,\sigma)\mu\nu}&=\partial_\mu\phi\partial_\nu\phi +\partial_\mu\chi\partial_\nu\chi +\partial_\mu\sigma\partial_\nu\sigma \\ &-\frac{1}{2}g_{\mu\nu}[\partial^\alpha\phi\partial_\alpha\phi +\partial^\alpha\chi\partial_\alpha\chi +\partial^\alpha\sigma\partial_\alpha\sigma +V(\phi ,\chi ,\sigma)].\label{4.20}
  \end{split}
 \end{equation}
Using the spherical symmetric line element given in equation (\ref{3.31}), we follow Refs.~\cite{E.S,W.M,M. Al,Balakrishna} and write Einstein's equations as:\\
\textit{the $t-t$ component}
\begin{equation}
\begin{split}
(g^2)'&= -g^2\left(\frac{g^2-1}{r}\right)+4\pi Gg^2r\left[\frac{g^2}{N^2}\dot{\sigma^2}+\sigma'^2 +\phi'^2 + \chi'^2\right]\\& +4\pi Gg^2r\left[g^2\left(\frac{1}{2}\sigma^2(\sigma-\sigma_0)^2 +(\phi^2 +\chi^2)^2 -\phi (\phi^2 -3\chi^2)+[(\sigma-1/2\sigma_0)^2-9/4](\phi^2 +\chi^2)\right)\right],\label{4.24}
\end{split}
\end{equation}
\textit{the $r-r$ component}
\begin{equation}
\begin{split}
(N^2)'&= N^2\left(\frac{g^2-1}{r}\right)+4\pi Gr\left[g^2\dot{\sigma^2}+N^2(\sigma'^2 +\phi'^2 + \chi'^2)\right]\\& +4\pi Gr\left[N^2g^2\left(\frac{1}{2}\sigma^2(\sigma-\sigma_0)^2 +(\phi^2 +\chi^2)^2 -\phi (\phi^2 -3\chi^2)+[(\sigma-1/2\sigma_0)^2-9/4](\phi^2 +\chi^2)\right)\right],\label{4.25}
\end{split}
\end{equation}
\textit{the $t-r$ component}
\begin{eqnarray}
\dot{g}=8\pi Grg\dot{\sigma}\sigma'.\label{4.26}
\end{eqnarray}


In order to find the Klein-Gordon equations we use Eqs.~(\ref{kg46})-(\ref{kg48}) with the potential defined in equation (\ref{4.15}). They can be written as
\begin{eqnarray}
\ddot{\sigma}-\frac{\dot{(N^2)}\dot{\sigma}}{2N}  =\frac{(N^2)'\sigma'}{2g^2}+\frac{N^2}{g^2}\left[\sigma'' -\frac{(g^2)'}{2g^2}\sigma' -\frac{\dot{(g^2)}}{N^2}\dot{\sigma}\right]+\frac{2N^2\sigma}{rg^2}  -N^2V_{\sigma},\label{4.16}
\end{eqnarray}
 \begin{eqnarray} 
 \phi''=\frac{(g^2)'}{2g^2}\phi' -\frac{(N^2)'}{2N^2}\phi' -\frac{2\phi'}{r} + g^2V_{\phi},\label{4.17}
 \end{eqnarray}
 \begin{eqnarray} 
 \chi''=\frac{(g^2)'}{2g^2}\chi' -\frac{(N^2)'}{2N^2}\chi' -\frac{2\chi'}{r}+ g^2V_{\chi},\label{4.18}
\end{eqnarray}
where $V_{\sigma}=\frac{\partial V(\sigma,\phi ,\chi)}{\partial\sigma}$, $V_{\phi}=\frac{\partial V(\sigma,\phi ,\chi)}{\partial\phi}$, and $V_{\chi}=\frac{\partial V(\sigma,\phi ,\chi)}{\partial\chi}$.

As done in Sec.~\ref{sec1}, we consider change of variables ($A(r,t)=g^2$, $C(r,t)=[g(r,t)/N(r,t)]^2$) in other to deal with the non-linearity present in the Einstein's equation. It is also convenient to perform suitable re-scaling of the parameters leading to dimensionless quantities. For this reason, we use the following dimensionless variables
$4\pi G = 1$, $r\rightarrow r/m_\sigma$,  $C\rightarrow Cm_\sigma^2/\omega^2$ and $t\rightarrow \omega t$. Here $m_\sigma$ is the mass of the $\sigma$ field computed at the true vacuum i.e., $m_\sigma^2=V_{\sigma\sigma}|_{\sigma=0}=\sigma_0^2$.
  
  
The coupled Einstein-Klein-Gordon equations now take the form
\begin{equation}
\begin{split}
A'&=A\left(\frac{1-A}{r}\right)+ Ar\left[C\dot{\mathbf{\sigma}^2}+\mathbf{\sigma}'^2 +\phi'^2 +\mathbf{\chi}'^2 \right]\\&+ A^2r\left[\frac{1}{2}\sigma^2(\sigma-\sigma_0)^2 +(\phi^2 +\chi^2)^2 -\phi (\phi^2 -3\chi^2)+[(\sigma-1/2\sigma_0)^2-9/4](\phi^2 +\chi^2)\right],\label{ekg1}
\end{split}
\end{equation}
\begin{equation}
\begin{split}
C'&=\frac{2C}{r} - \frac{2CA}{r}\\ &+2CAr\left[\frac{1}{2}\sigma^2(\sigma-\sigma_0)^2 +(\phi^2 +\chi^2)^2 -\phi (\phi^2 -3\chi^2)+[(\sigma-1/2\sigma_0)^2-9/4](\phi^2 +\chi^2)\right],
\end{split}
\end{equation}
 \begin{eqnarray}
 C\ddot{\mathbf{\sigma}}=-\frac{1}{2}\dot{C}\dot{\mathbf{\sigma}}+\mathbf{\sigma}'' +\mathbf{\sigma}'\left(\frac{2}{r}-\frac{C'}{2C}\right)-AV_{\sigma},
 \end{eqnarray}
 \begin{eqnarray} 
 C\phi''=\frac{C'\phi'}{2}-\frac{2\phi'C}{r}+CA V_{\phi},
 \end{eqnarray}
 \begin{eqnarray} 
 C\mathbf{\chi}''=\frac{C'\mathbf{\chi}'}{2}-\frac{2\mathbf{\chi}' C}{r}+CA V_{\chi},\label{ekg2}
 \end{eqnarray}
 \begin{eqnarray}
 \dot{A}=rA\dot{\mathbf{\sigma}}\mathbf{\sigma}'.
 \end{eqnarray}
 
The simplest solutions to equations (\ref{ekg1})-(\ref{ekg2}) are periodic expansions of the form
\begin{eqnarray}
 A(r,t)=\displaystyle\sum_{j= 0}^{J_{max}} A_{2j}(r)\cos(2j\omega t),
 \end{eqnarray}
\begin{eqnarray}
C(r,t)=\displaystyle\sum_{j= 0}^{J_{max}} C_{2j}(r)\cos(2j\omega t),
\end{eqnarray}
\begin{eqnarray}
\mathbf{\sigma}(r,t)=\displaystyle\sum_{j= 1}^{J_{max}}\sigma_{2j-1}(r)\cos[(2j-1)\omega t],
\end{eqnarray}
where $\omega$ is the fundamental frequency and again will be absorbed in the time variable. $J_{max}$ is the value of $j$ at which the series is truncated for numerical computation.
 
 \subsection{Numerical analysis}
 We want to study the effect of network of domain walls with an oscillating soliton star, with the help of a Lagrangian containing three scalar fields ($\sigma, \phi, \chi$), where the sigma field ($\sigma$) serves as shell of the star  (or the host field) and the remaining two scalar fields ($\phi$ and $\chi$) are responsible for the formation of the network. In other to accomplish this task it is necessary to consider two cases where we use two types of perturbations.



\subsubsection{Case 1. Perturbations of the potential}
\label{sec31}

In this case we study the formation of network of domain walls with oscillating soliton stars, by changing $\sigma$ to work around the parameter $\sigma_0$. This leads to small perturbations of the potential, resulting in the possible disturbance of the network around the surface of the star. This is done by considering $\sigma\approx\sigma_0 +\eta$, where $\eta\ll\sigma_0$ ($\eta$ is a perturbation field which is placed in Eqs.~(\ref{ekg1})-(\ref{ekg2}) that after performing suitable computations is replaced back in favor of $\sigma$). Depending on how we choose the relation between $\sigma$ and $\sigma_0$ based on equation (\ref{4.11}) will tell us whether the network is entrapped inside the star, or the network is exactly on its surface, or outside the star.
Thus leading to these three possibilities:\\
a) The network forming inside the star ($\sigma\approx  \sigma_0+\eta$)\\
b) The network forming on the surface of the star ($\sigma\approx (1/2) \sigma_0+\eta$)\\
c) The network forming outside the star ($\sigma\approx (1/4)\sigma_0+\eta$).\

By  imposing these conditions (a, b, and c) on equations  (\ref{ekg1})-(\ref{ekg2}), we find sets of EKG equations for each possibility under this kind of perturbations. We solve the EKG equations by invoking the same boundary conditions used in section (\ref{sec22}) only that in this case we are working with three scalar fields. The series is truncated after $j_{max}=2$ and the Fourier coefficient are set to zero. 
 \begin{figure}[h!]
\centering
\subfigure[Network inside the oscillaton]{
\includegraphics[width=.4\textwidth]{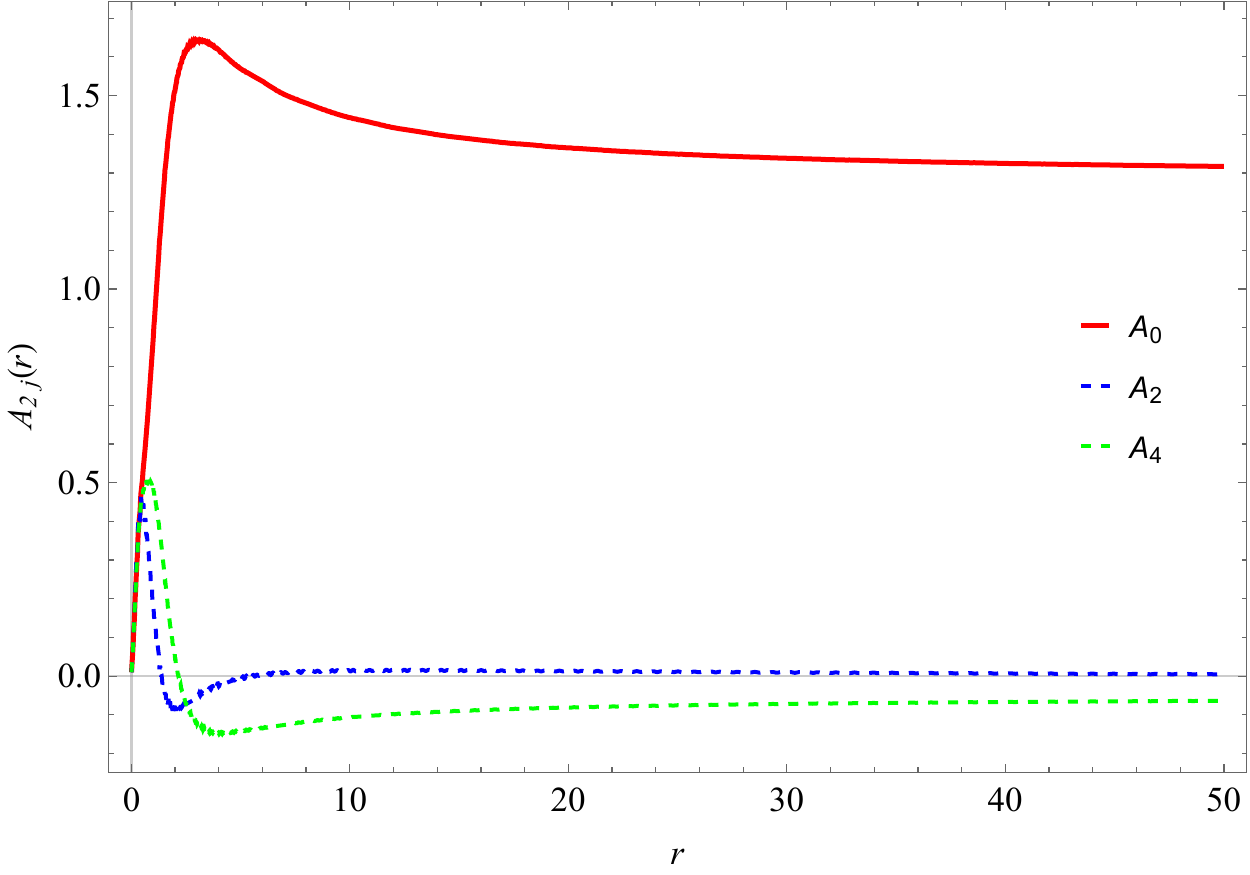}
}
\subfigure[Network on the oscillaton surface]{
\includegraphics[width=.4\textwidth]{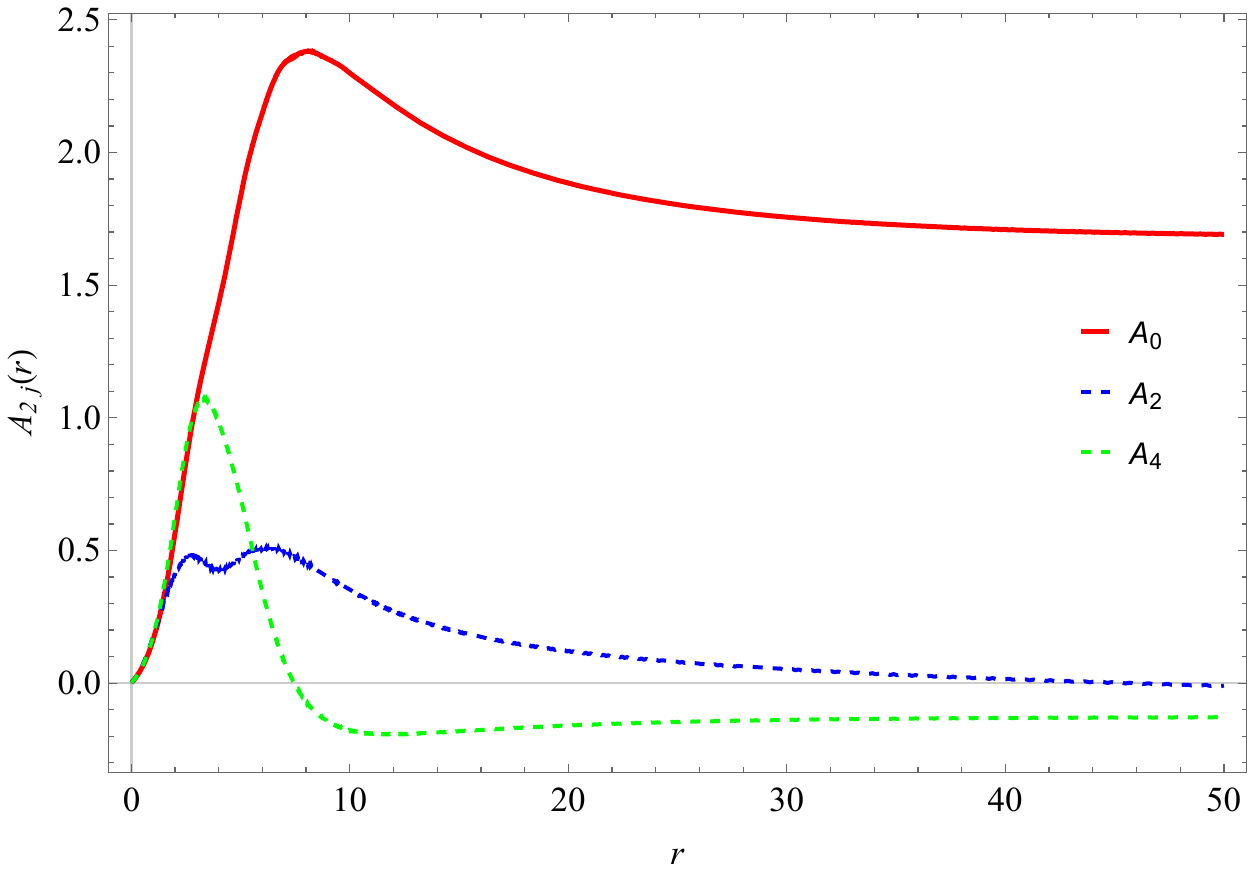}
}
\subfigure[Network outside the oscillaton]{
\includegraphics[width=.4\textwidth]{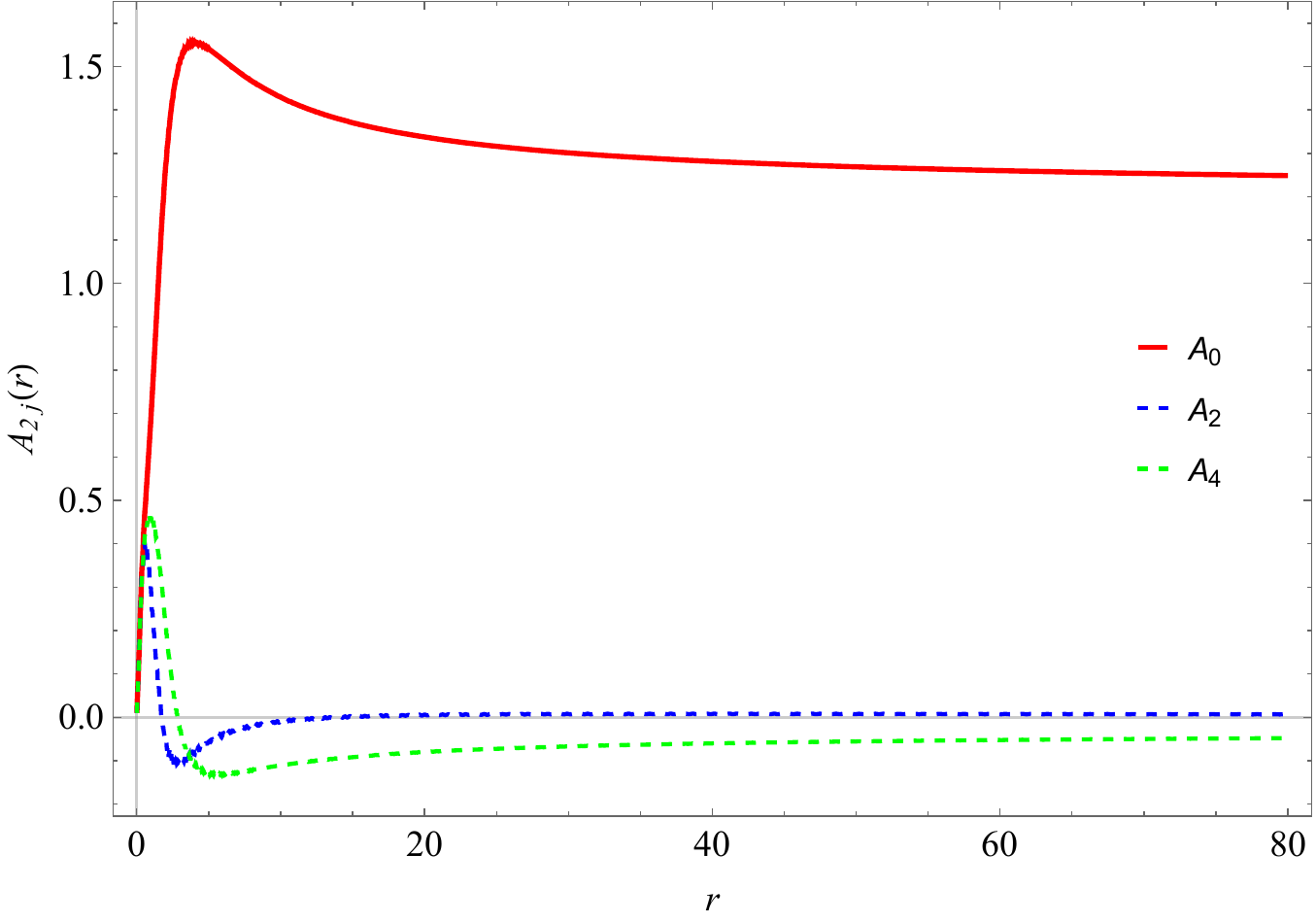}
}
\caption{A typical solution to the truncated eigenvalue equations ($j_{max}=2$) of the expansion of the metric quantity $A(r,t)$ for the three possibilities under this perturbations. These simulations are computed for the central densities $\sigma_1(0)\approx0.1$ (surface), $\sigma_1(0)\approx0.2$ (other cases), $\phi_1(0)\approx0$, and $\chi_1(0)\approx0$.}
\label{figs3}
\end{figure}
The Figs.~\ref{figs3} demonstrate typical numerical results for the metric $g_{rr}=A(r,t)$. The convergence of the series is significant, regardless of the non-linearity of the EKG equations.
\begin{figure}[h!]
\centering
\subfigure[ Oscillaton with network  inside ]{
\includegraphics[width=.4\textwidth]{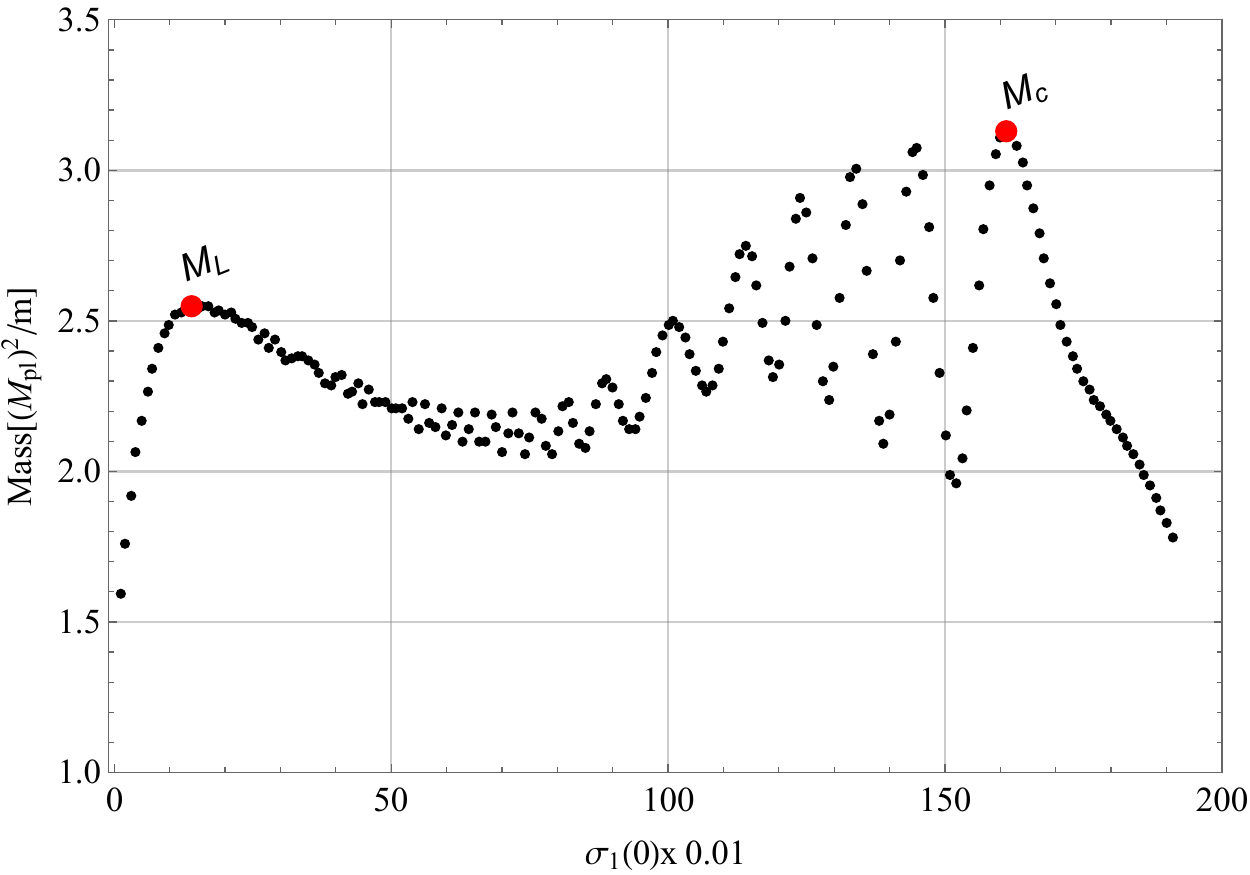}
}
\subfigure[ Oscillaton with network  on the surface]{
\includegraphics[width=.4\textwidth]{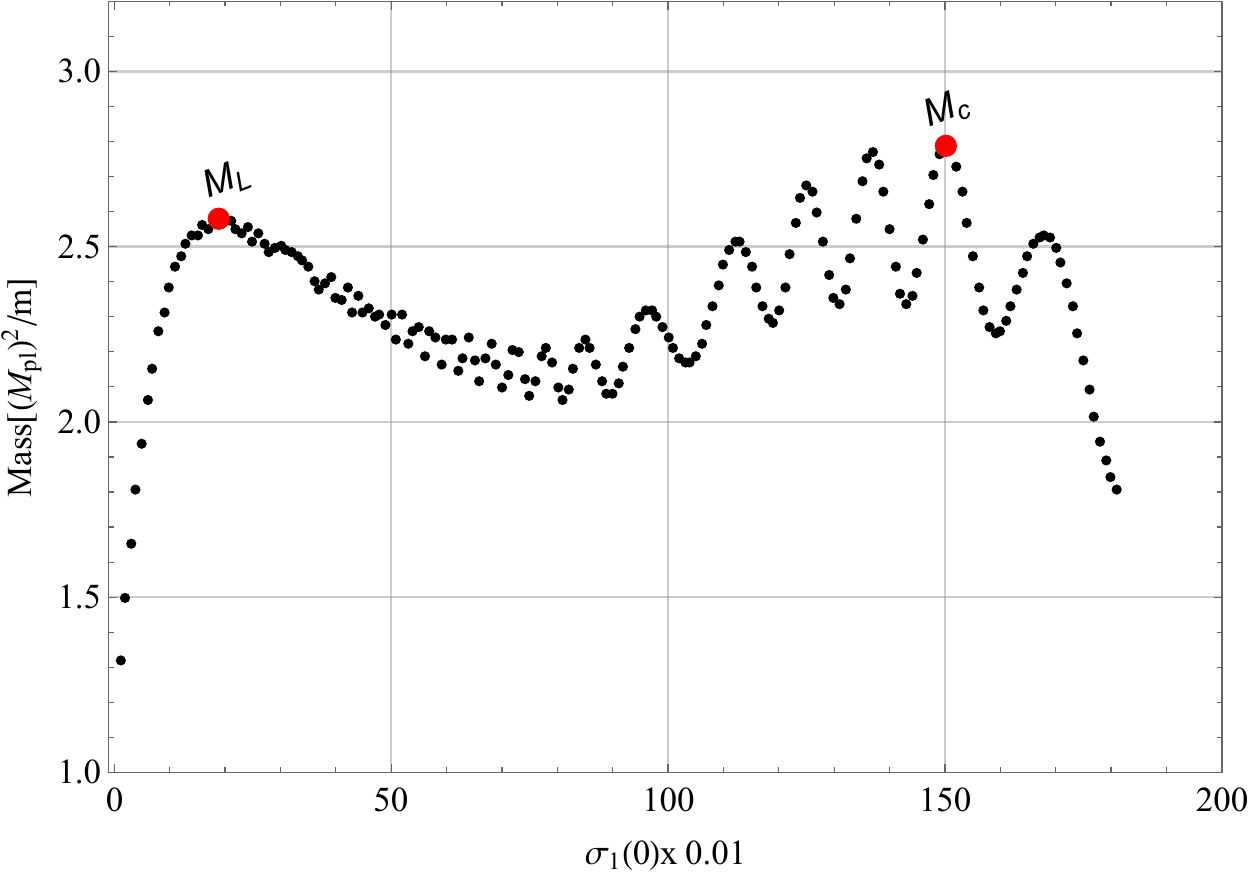}
}
\subfigure[Oscillaton with network  outside ]{
\includegraphics[width=.4\textwidth]{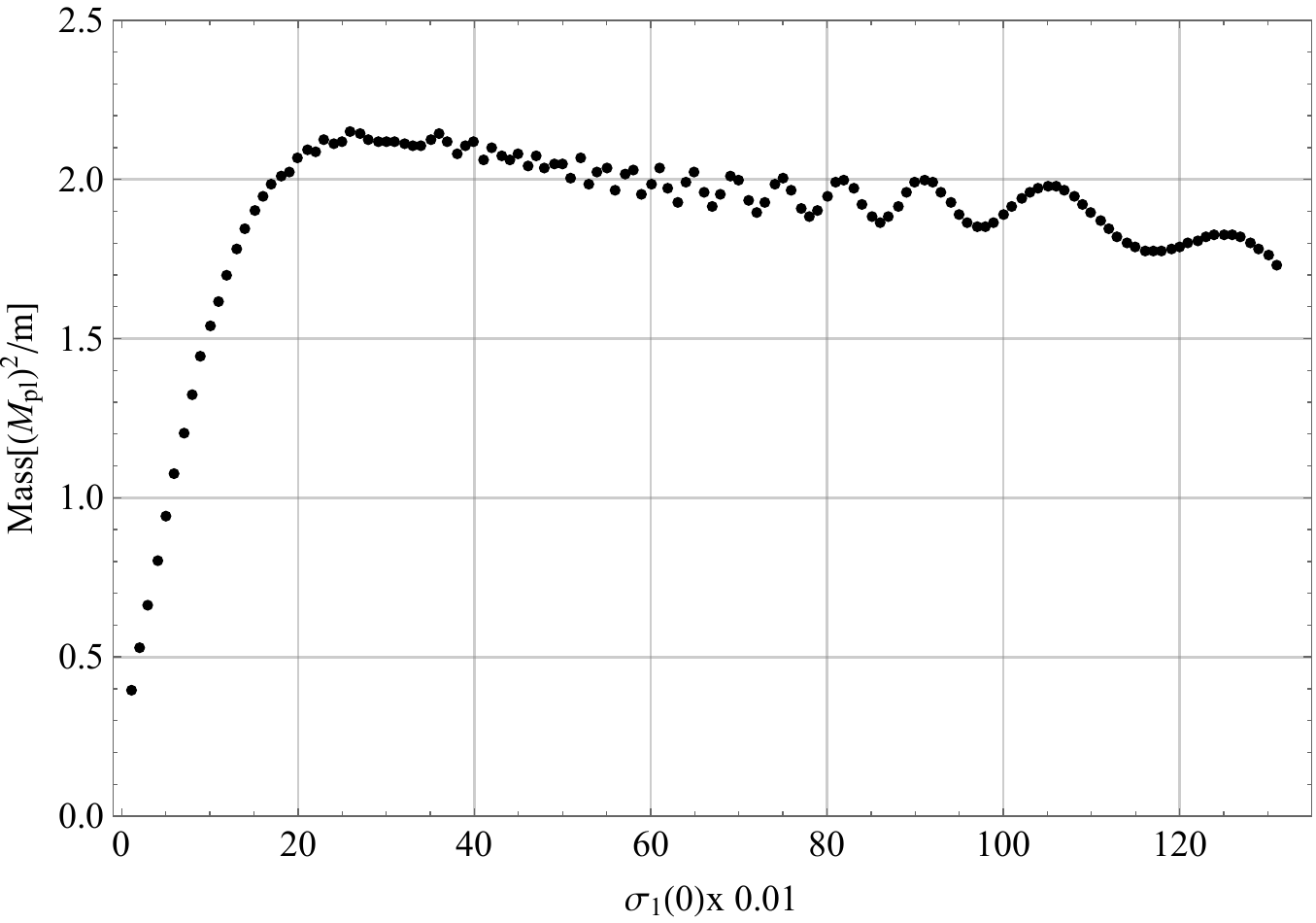}
}
\caption{The total mass $M$ of the oscillating soliton star (in units of $M^2_{Planck}/m_\sigma$) is plotted as a function of central density $\sigma_1(0)$, for all three possibilities. The central densities for the other two fields are fixed at $\phi_1(0)\approx0$ and $\chi_1(0)\approx0$. The circles represent actual configurations resulting from solutions to the eigenvalue equations.}
\label{figs4}
\end{figure}
 
Let us study the dynamics of the mass profiles of the oscillaton stars under this perturbation. In Fig.~\ref{figs4}(a) { an important point arises, we observed that, the mass increases monotonically until it reaches a point (local maximum $M_L$) $M_L\approx 2.55M_{Planck}^2/m_\sigma$, after which it starts to increase again to a point $M_c\approx 3.15M_{Planck}^2/m_\sigma$ (the critical mass). After this the mass drops, oscillates and finally collapses. This presents a different behavior from the mass profile well-known in other theories. This exhibits a behavior which we call ``the bouncing stability". We can explain this behavior as the shifting of the critical mass before collapse further. Meaning the star have the ability to accrue more mass before decaying into a black hole, and this leads to the increment of the lifespan of the star. Indeed, we can associate this feature to the network of domain walls present inside the surface of the star}.
 
{The behavior of mass of the oscillaton with network on its surface  is shown in Fig.~\ref{figs4}(b), which is similar to the one with network inside with $M_L\approx 2.58M^2_{Planck}/m_\sigma$ and $M_c\approx 2.79M_{Planck}^2/m_\sigma$. The location of the maximum shifts to higher values of $\sigma_1(0)$, clearly the network on the surface has affected the mass profile}.
 
{Looking at the behavior shown in Fig.~\ref{figs4}(c), the mass increases monotonically until it reaches a value $M_L\approx 2.12M^2_{Planck}/m_\sigma$. After this it decreases. It physically makes sense because in this simulation, the network is outside the star, which is the same as saying there is no network on the surface, or inside the star, hence resulting in this behavior}.
 
\subsubsection{Case 2. Perturbations of the shell of the star }

The focus of this case is to study small perturbations applied to  the `surface term' $(\sigma-\frac{1}{2}\sigma_0)^2$ in the potential and to investigate the effect that these perturbations will have on the mass profile of the oscillaton. We perturb the shell by the introduction of a perturbation parameter $\lambda$ into the surface term in the potential and fix the parameter $\sigma_0\approx 1$, leading to the potential given by
\begin{eqnarray}
  V=\frac{1}{2}\sigma^2(\sigma -1)^2+(\phi^2 +\chi^2)^2-\phi(\phi^2 -3\chi^2)+[(\sigma -\frac{1}{2}\lambda)^2 -\frac{9}{4}](\phi^2 +\chi^2).\label{4.8}
\end{eqnarray}
Carefully adjusting $\lambda$ around $\sigma_0$, produces small perturbations of shell around the network, by  slightly shifting the shell to entrap the network inside star, or leaving the network outside the star. Here when the shell coincide exactly with the surface of the network, the surface term in the potential vanishes i.e., $(\sigma-\frac{1}{2}\lambda)^2\approx 0$. 
This generates three possibilities:\\
a) The network forming inside the star (when we consider $\lambda\approx 2$),\\
b) The network forming on the surface of the star, i.e., $(\sigma-\frac{1}{2}\lambda)^2\approx 0$,\\
c) The network forming outside the star ($\lambda\approx \frac{1}{2}$).
  
In order to find solutions to the EKG equations for all the three possibilities under this case, we invoke the arguments presented above (a, b and c) on the potential. These conditions serve as the source of the perturbations leading to the said possibilities. Following the usual numerical routines, we extract and discuss the resulting graphs. 

\begin{figure}[h!]
\centering
\subfigure[Network of domain walls inside the oscillaton]{
\includegraphics[width=.4\textwidth]{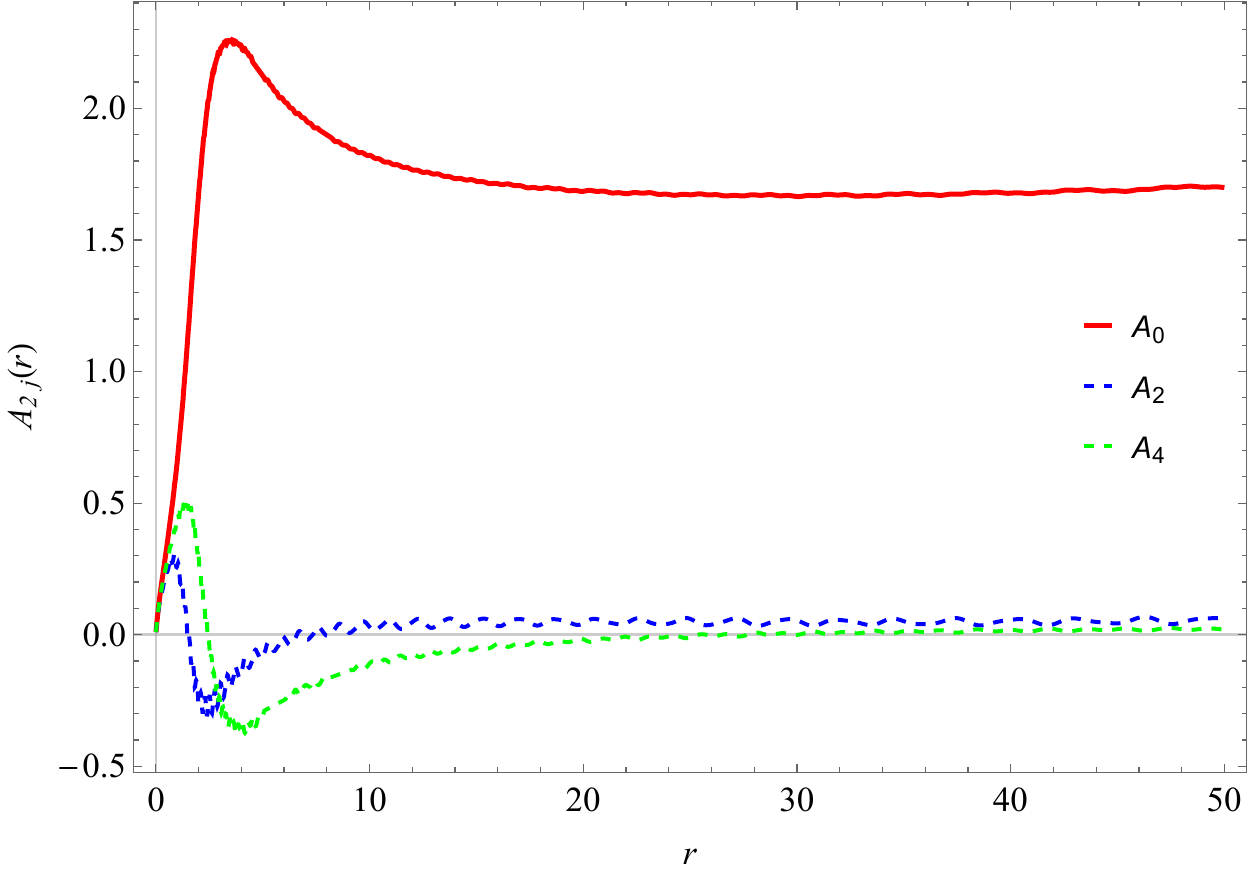}
}
\subfigure[Network of domain walls on the oscillaton surface]{
\includegraphics[width=.4\textwidth]{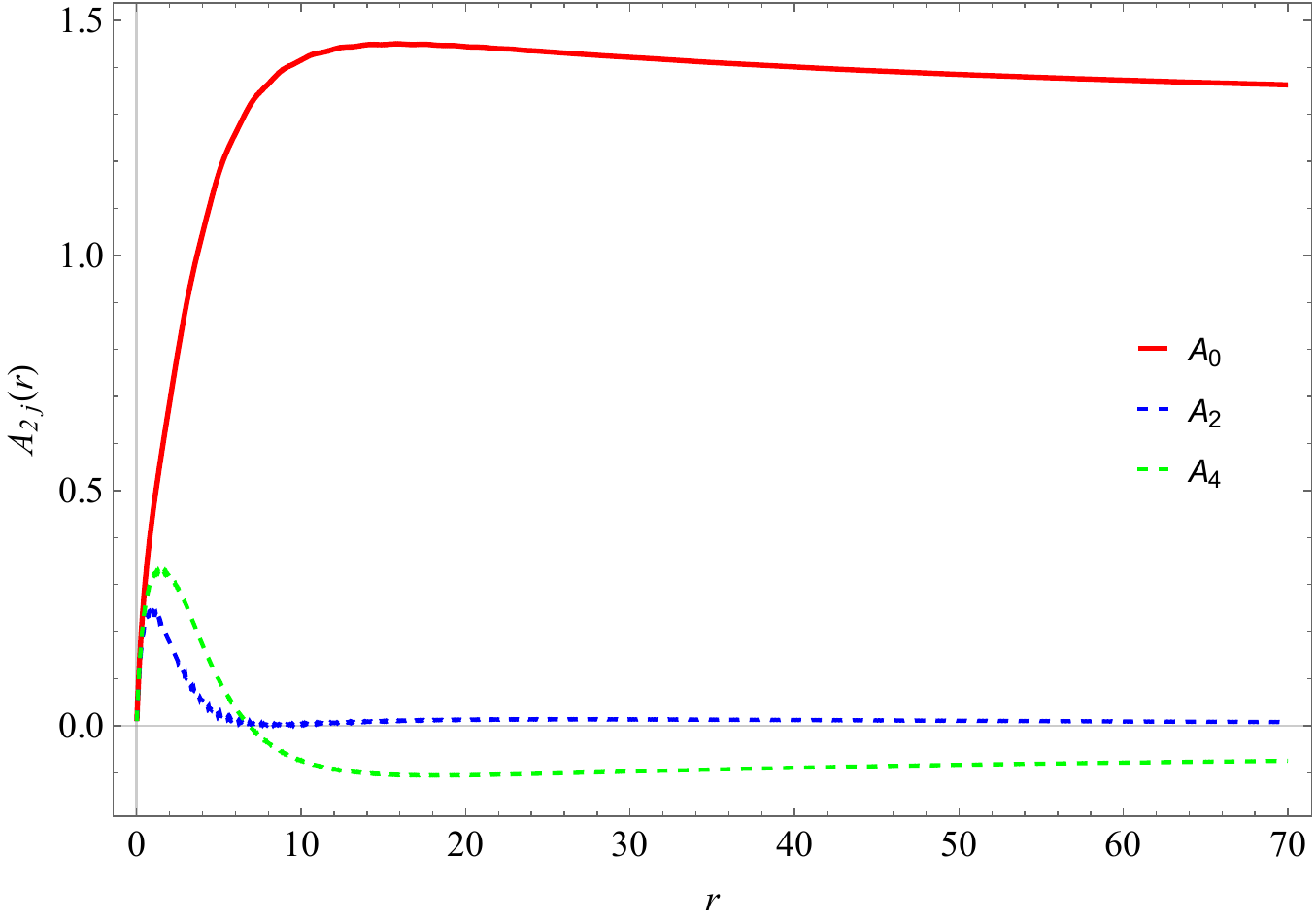}
}
\subfigure[Network of domain walls outside the oscillaton]{
\includegraphics[width=.4\textwidth]{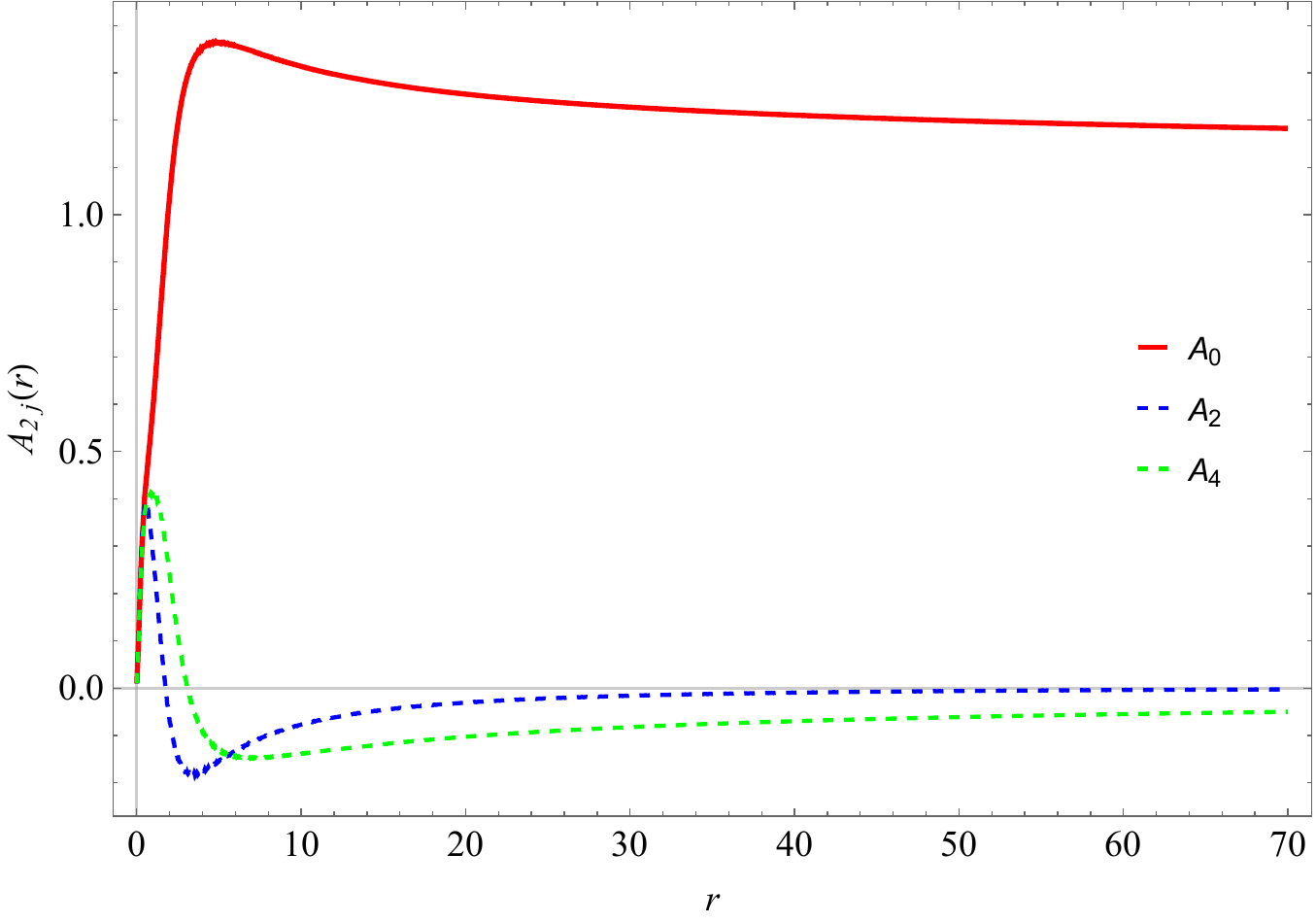}
}
\caption{A typical solution to the truncated eigenvalue equations ($j_{max}=2$) of the expansion of the metric $A(r,t)$ for the three possibilities under this perturbations. These simulations are computed for the central densities $\sigma_1(0)\approx0.2$(outside), $\sigma_1(0)\approx0.1$(other cases), $\phi_1(0)\approx0$ and $\chi_1(0)\approx0$.}
\label{figs5}
\end{figure}

We can see from Fig.~\ref{figs5}, a typical radial metric function $A(r,t)$, for the case of central density of $\sigma_1(0)\approx0.2$(outside) and $\sigma_1(0)\approx0.1$(inside and surface) is plotted. The solid line shows the first term of the Fourier series expansion, while the dashed lines are the second and third terms. This rapid convergence of the series is typical of all the configurations we have calculated.
\begin{figure}[h!]
\centering
\subfigure[ Oscillaton with network of domain walls inside ]{
\includegraphics[width=.4\textwidth]{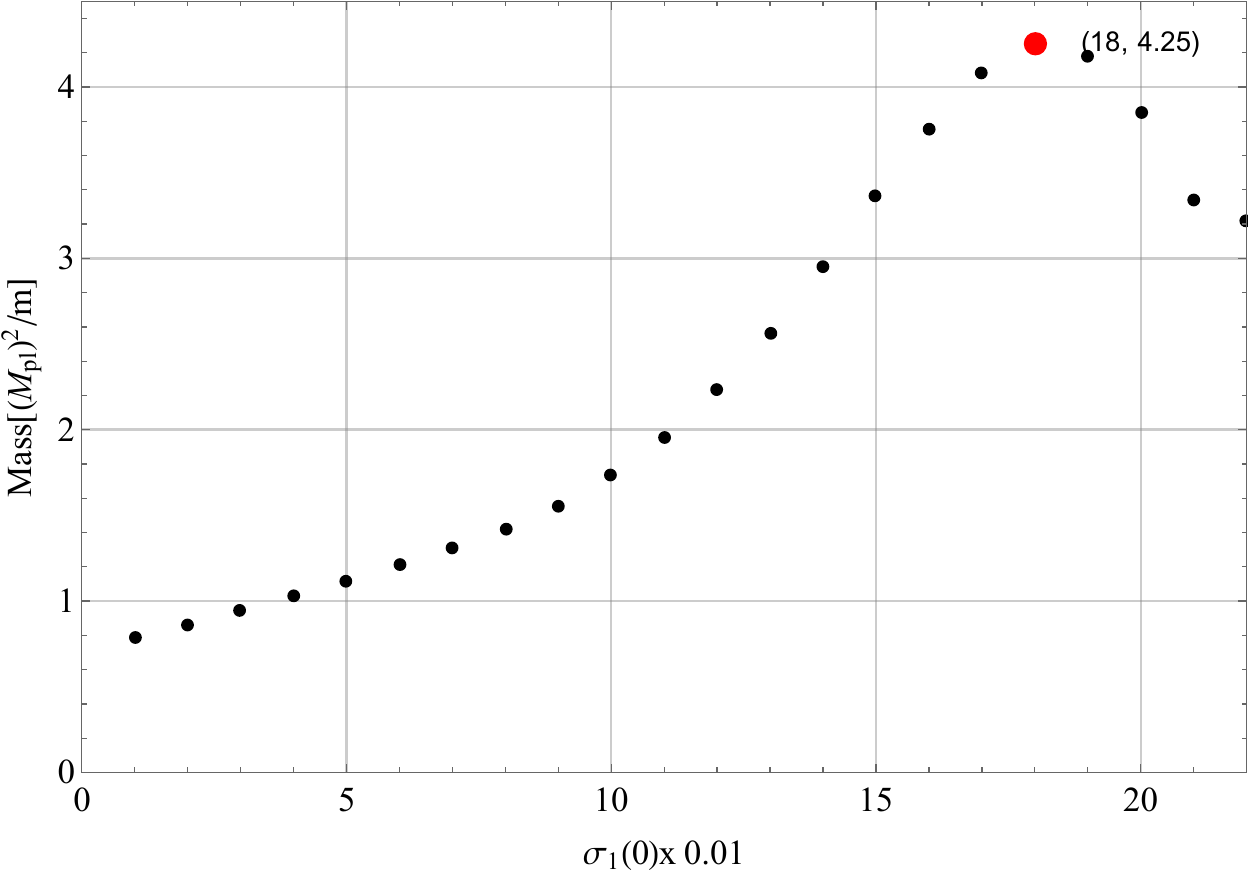}
}
\subfigure[ Oscillaton with network of domain walls on the surface]{
\includegraphics[width=.4\textwidth]{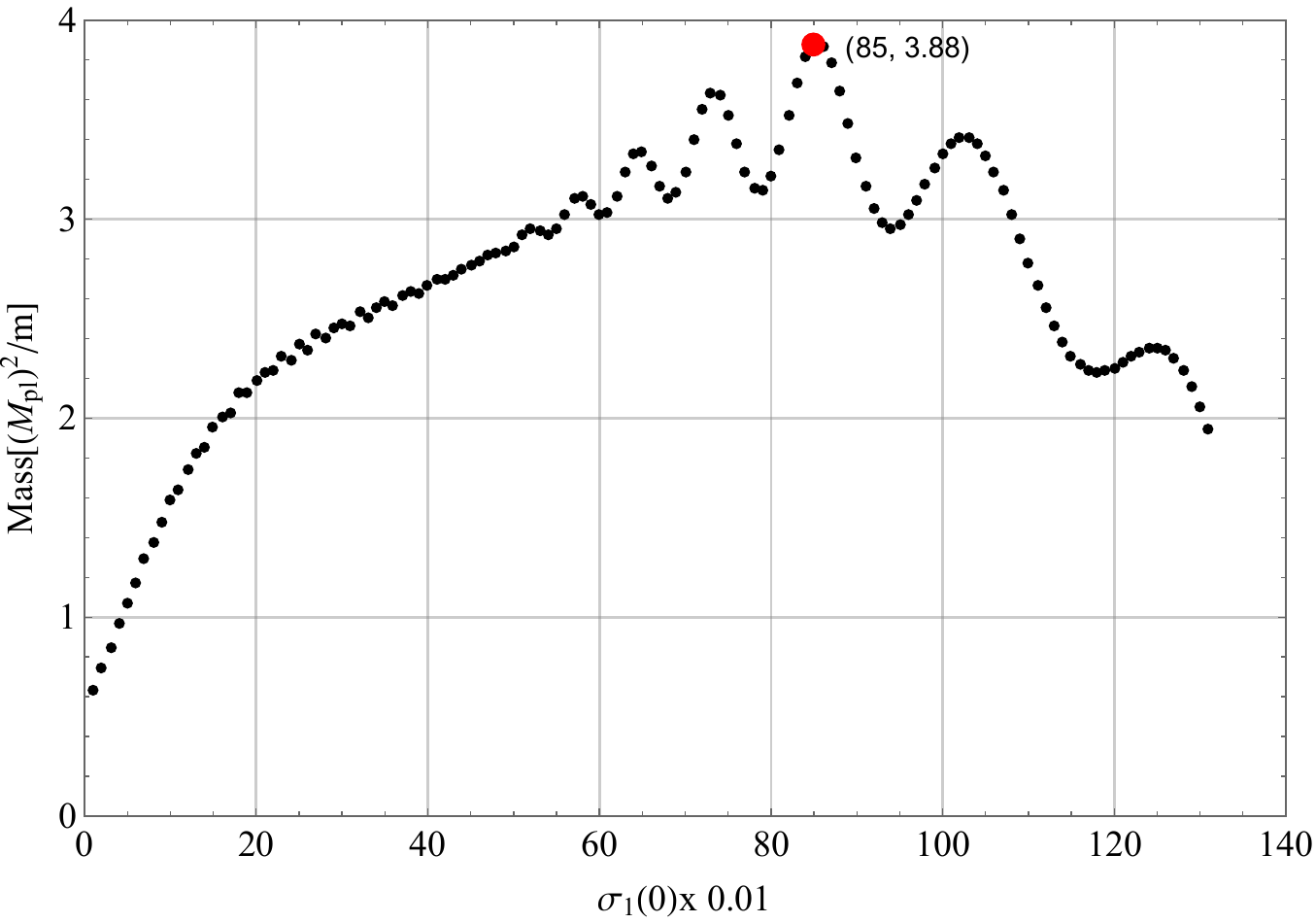}
}
\subfigure[ Oscillaton with network of domain walls outside]{
\includegraphics[width=.4\textwidth]{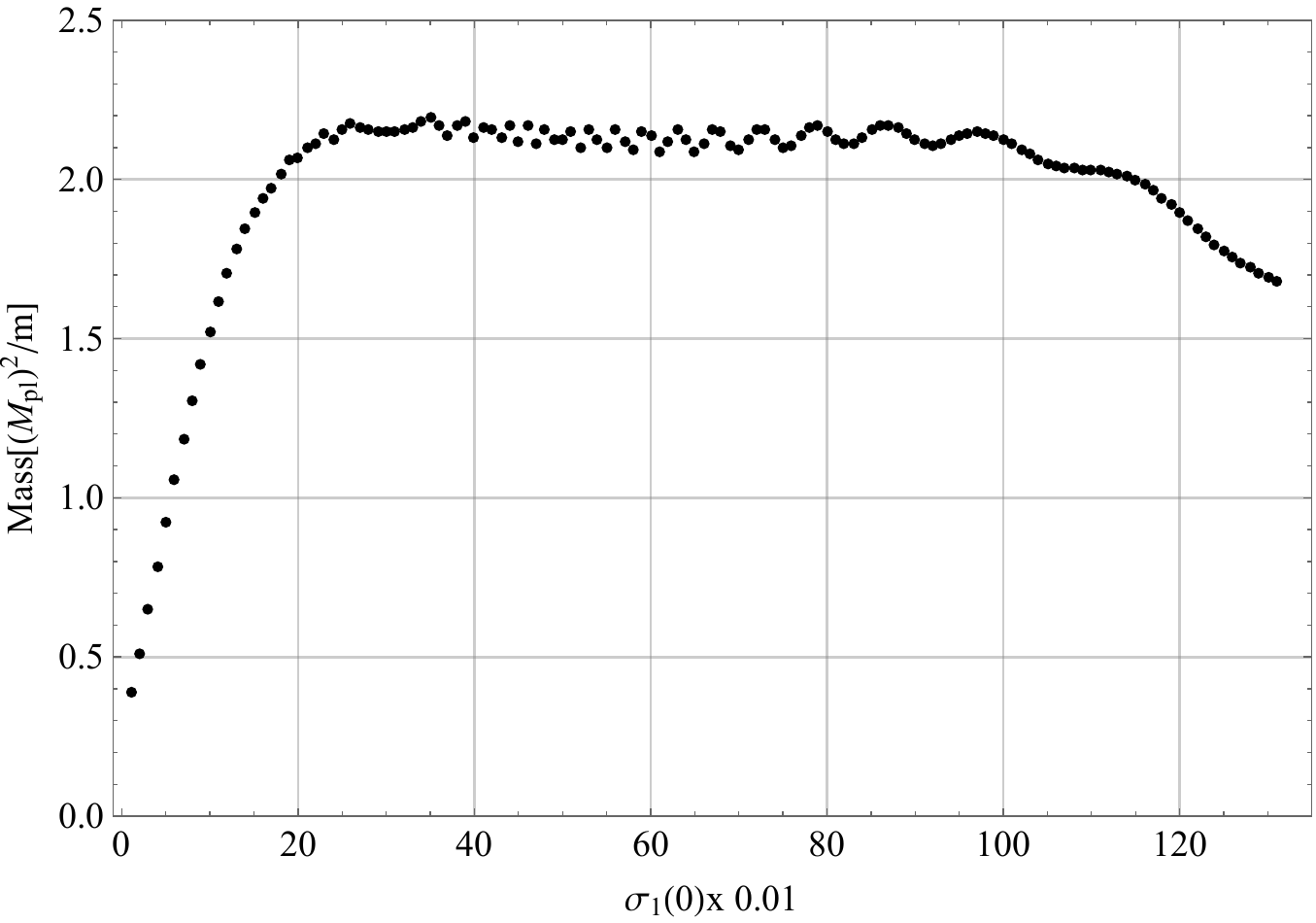}
}
\caption{The total mass $M$ of the oscillating soliton star (in units of $M^2_{Planck}/m_\sigma$) is plotted as a function of central density $\sigma_1(0)$, for all three possibilities. The central densities for the other two fields are fixed at $\phi_1(0)\approx0$ and $\chi_1(0)\approx0$. The circles represent actual configurations resulting from solutions to the eigenvalue equations.}
\label{figs6}
\end{figure}

The mass profile from the numerical results for different oscilatons as function of $\sigma_1(0)$ are shown in Fig.~\ref{figs6}. In Fig.~\ref{figs6}(a) we have the mass profile for the simulation when the network is sitting inside the oscillaton. This case reveals a different behavior as compared to that of first case. Here we see that the mass profile exhibit again ``the bouncing stability" with a local  maximum $M_L\approx4.25M^2_{Planck}/m_\sigma$ and a critical central density of $\sigma_c\approx0.18$.
 
The mass curve shown in Fig.~\ref{figs6}(b), is similar to the configuration presented in the first case, but with the presence of a maximum mass $M_c\approx 3.88 M_{Planck}^2/m_\sigma$. It indicates that, again the star is prevented from decaying into a black hole due to the presence of network on its surface. We find that in both perturbations it is convenient to have the network on the surface of the star if we want to ensure the stability of the star.\ 
  
The  Fig.~\ref{figs6}(c), shows a general behavior similar to the mass profile in the first case with a maximum mass $M_c\simeq 2.18 M_{Planck}^2/m_\sigma$ at $\sigma_c(0)\simeq 0.26$. By comparison, this case present an oscillaton with a delayed decay.
  
\section{Discussions and Conclusions}
\label{sec3}

In this work we presented a model that admits formation of network of domain walls living on oscillatons. This was done by introducing a Lagrangian that contains three scalar fields that is coupled among themselves by a suitable potential. The potential was chosen to provide the standard spherical oscillating soliton star with a network of domain walls living on \cite{D.F}. 

 We used two types of perturbations to simulate the three possibilities; having the network of domain walls inside the star, or the network forming on the surface of the star, or the network sitting outside the star. These three scenarios can be thought of as three possible different experiences that an oscillaton can go through in the cosmological evolution. Sec.~\ref{sec31} presented the first case where we considered small perturbations to the whole potential of the system. This was made possible by working the host field $\sigma$ around $\sigma_0$ with the help of a perturbation field $\eta$. The next section presented the second case where we perturbed the surface term in the potential, by carefully adjusting a parameter $\lambda$ around $\sigma_0$. These simulations are challenging because of the non-linearity of the Einstein-Klein-Gordon equations and also due to dynamical characteristics of oscillating soliton stars, with no equilibrium configurations having static metric components. In this work we truncated the system of equations after a maximum of $j=j_{max}=2$. Although we have not proved analytically that the series represents an exact solution to the Einstein-Klein-Gordon equations for the three scalar field system, we have given strong evidence that the series indeed converges rapidly, in both cases (Figs.~\ref{figs3} and \ref{figs5}). 

Since the main objectives of this work is to investigate the effect of network of domain walls on the stability of the star, we particularly invested much effort in the mass profiling of all configurations in both perturbations. From this we have demonstrated that the network of domain walls indeed affects the stability of the star. Figs.~\ref{figs4}(b), \ref{figs6}(a), and \ref{figs6}(b) present a different behavior from the mass profiles well-known in other theories (for solitons and boson stars). We can say that the present object of study has two S-branches, the first S-branch is for the part of the star where there is no network and the second S-branch can be related to the network of domain walls on the star. From the simulations, we can say that it is energetically favorable for the network to be formed on the surface of the star, meaning, the formation of network on the surface of the star can be a cosmological mechanism that can ensure the stability of an oscillating soliton star. Also, we saw interesting results, when we simulated the possibility of having the network sitting outside the star --- Figs.~\ref{figs4}(c) and \ref{figs6}(c). We find out that their mass profiles are similar to that in usual theories. This physically makes sense, since network outside the star can be interpreted as no network on the star. 

To our knowledge this is the first time oscillating soliton stars with network of domain walls are studied. This investigation has led to the discovery of new type of self-gravitating objects and their existence could have significant cosmological and astrophysical implications. They can interplay the role with primordial black holes new cosmological scenarios via transition induced by perturbation of the network. As source of gravitational waves they can provide new signatures that may be detected in the form of gravitational and electromagnetic waves in the same event. When such objects are sufficiently close from each other, tidal forces can deform their surface in relation to the network destroying their stability driving them to collapse to black holes.\

Finally, noticed that in this work all the analysis were done without the consideration of temperature, i.e., we did not consider the effect of temperature on the network. The model presented gives room for the inclusion of temperature in the form of thermal corrections that will modify the Lagrangian. For instance, one might want to compute the critical temperature to form network due to QCD phase transition. This issue and many others will be addressed in the near future.

\acknowledgments

We would like to thank J.R.L. Santos and C.A.S. Silva for discussions. J.A.V. Campos would like to thank FAPESQ-PB/CNPq n$^0$ 77/2022 for financial support and F.A. Brito acknowledges CNPq and CNPq/PRONEX/FAPESQ-PB (Grant nos. 165/2018 and 309092/2022-1).
\newpage
\section{Appendix}

In this appendix, for the sake of the completeness, we add the relevant complementary graphs. 

\begin{figure}[!h]
\centering
\subfigure[{\bf \, Inside}-First case: The radial metric functions $C_0, C_2, C_4$. ]{
\includegraphics[width=.31\textwidth]{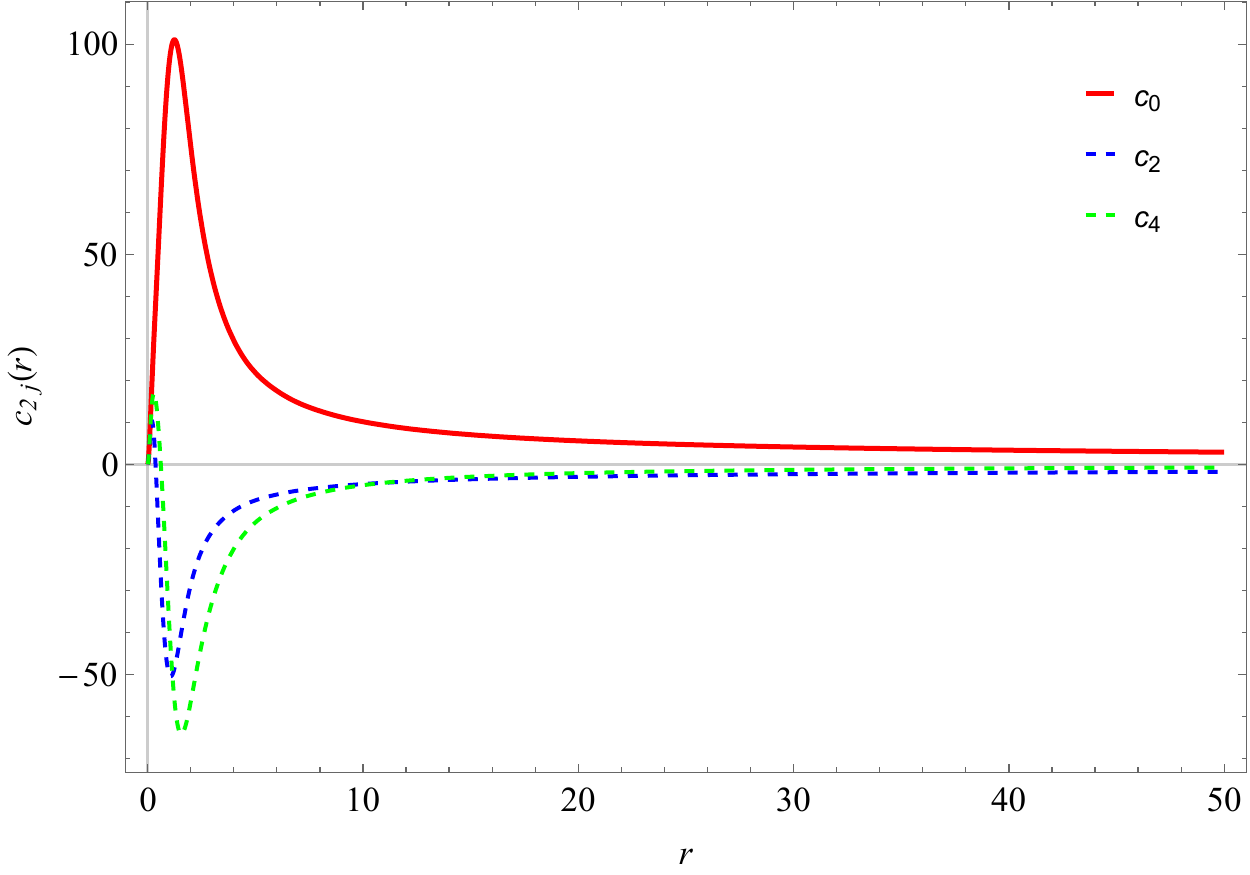}
}
\subfigure[{\bf \, Inside}-First case: The first two modes of the scalar field $\sigma$.]{
\includegraphics[width=.31\textwidth]{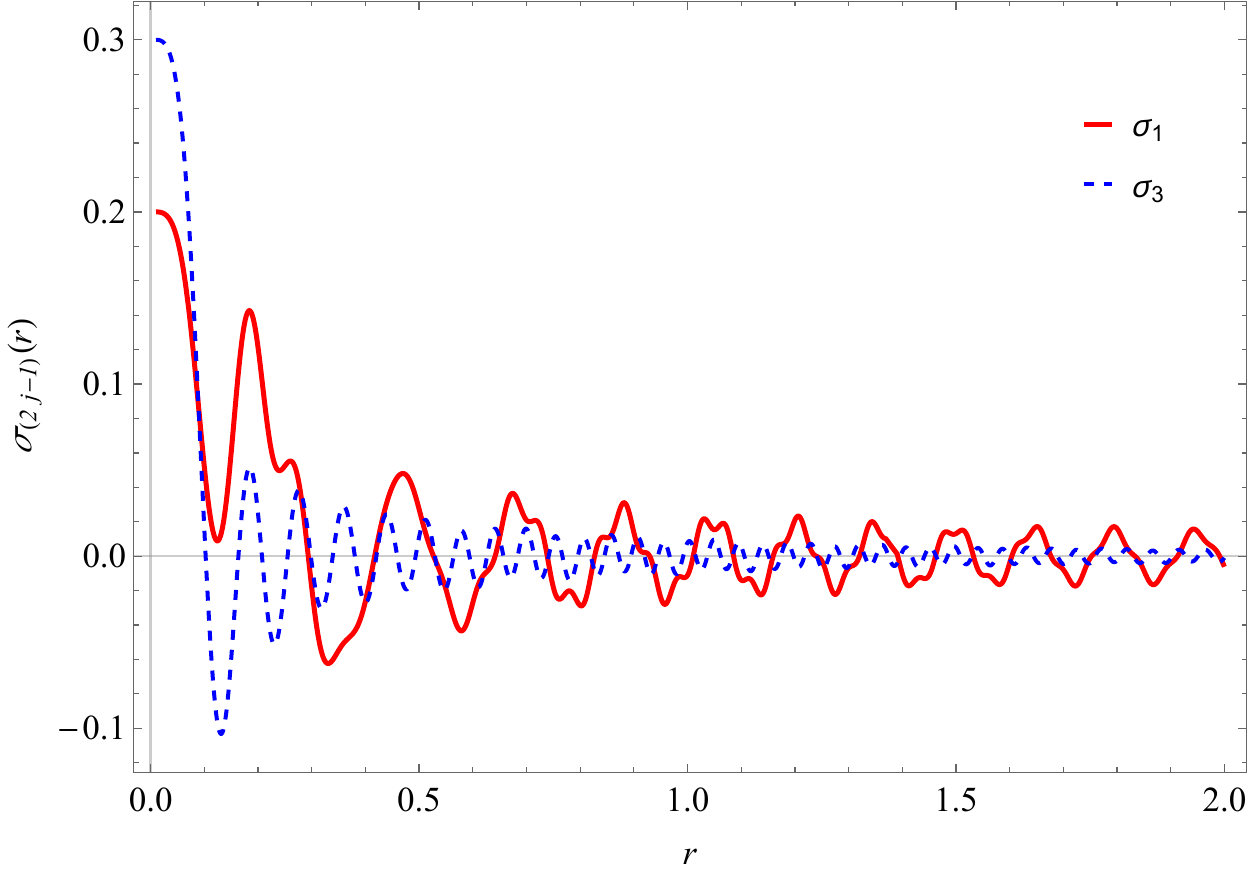}
}
\subfigure[{\bf \, Inside}-First case: The fields $\phi$ and $\chi$ inside the oscillaton.]{
\includegraphics[width=.31\textwidth]{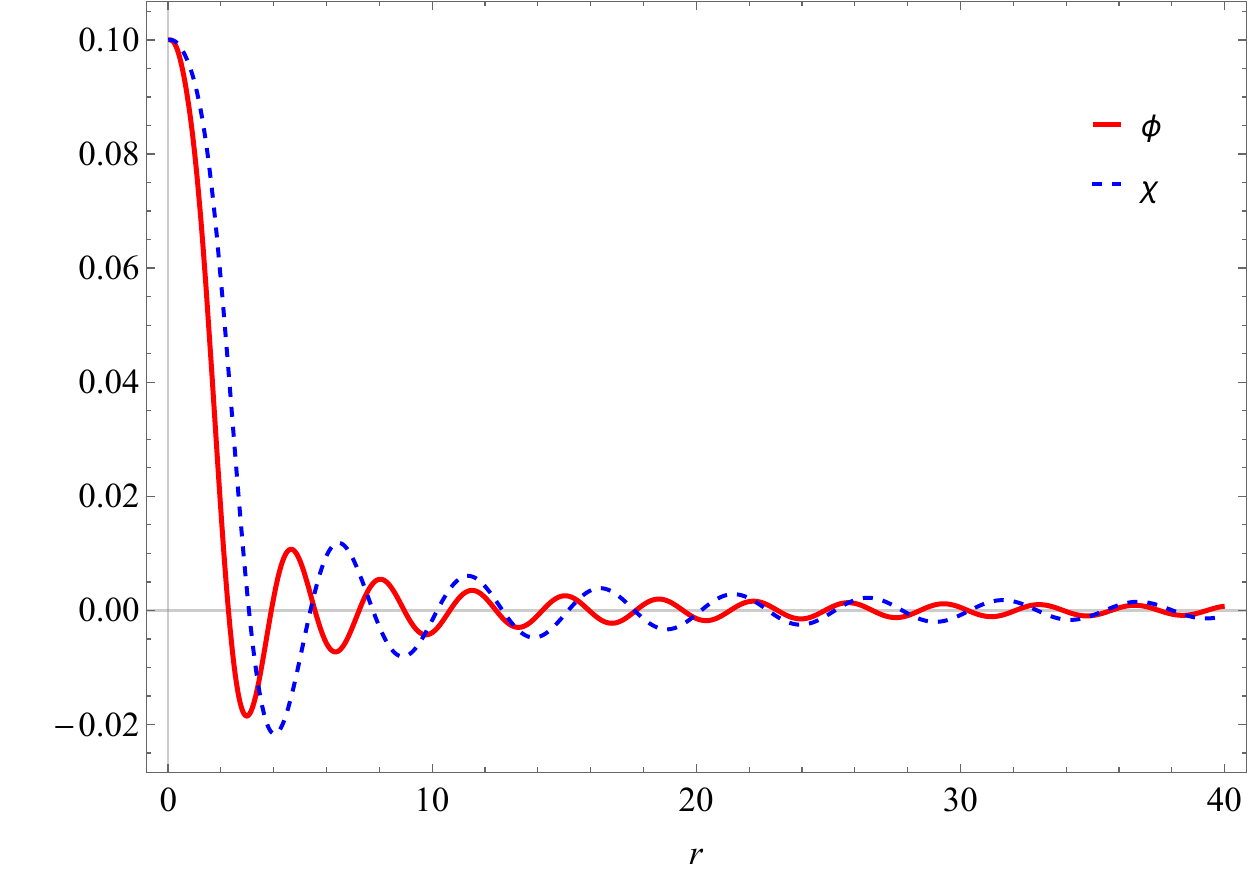}
}
\caption{These graphs are complementary of the Fig.~\ref{figs3} for the network around but inside the star.}
\label{figs-app-inside-1}
\end{figure}
\begin{figure}[h!]
\centering
\subfigure[{\bf \, Inside}-Second case: The radial metric functions $C_0, C_2, C_4$. ]{
\includegraphics[width=.31\textwidth]{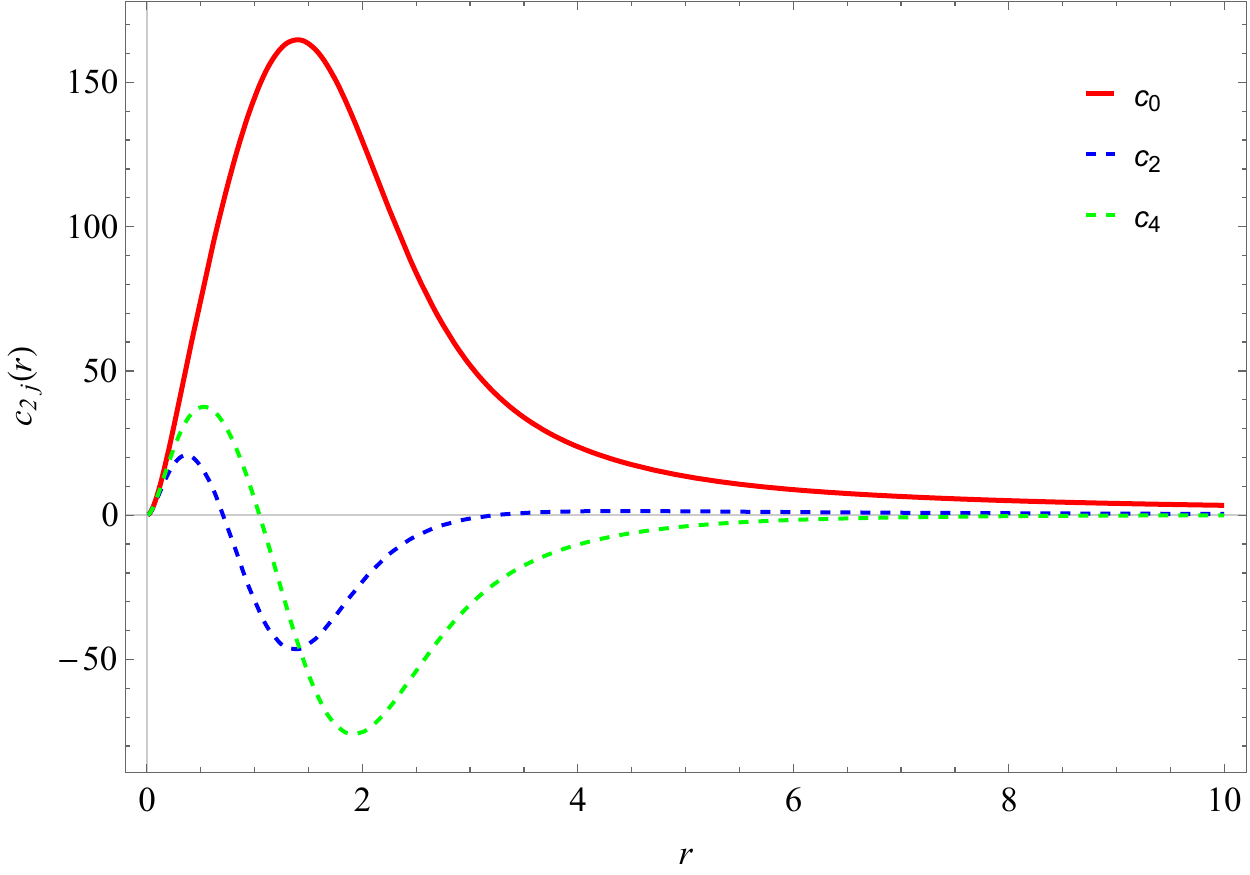}
}
\subfigure[{\bf \, Inside}-Second case: The first two modes of the scalar field $\sigma$.]{
\includegraphics[width=.31\textwidth]{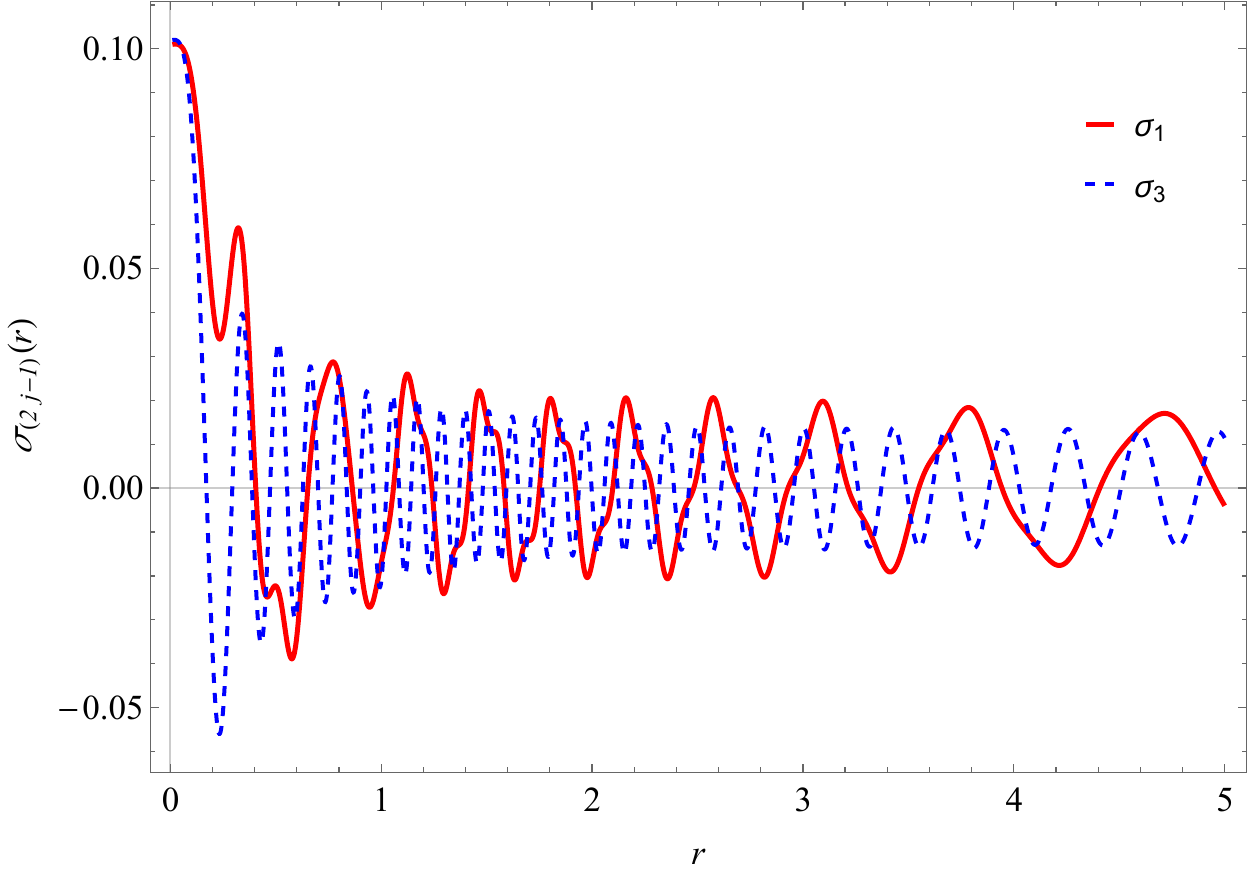}
}
\subfigure[{\bf \, Inside}-Second case: The fields $\phi$ and $\chi$ outside the oscillaton.]{
\includegraphics[width=.31\textwidth]{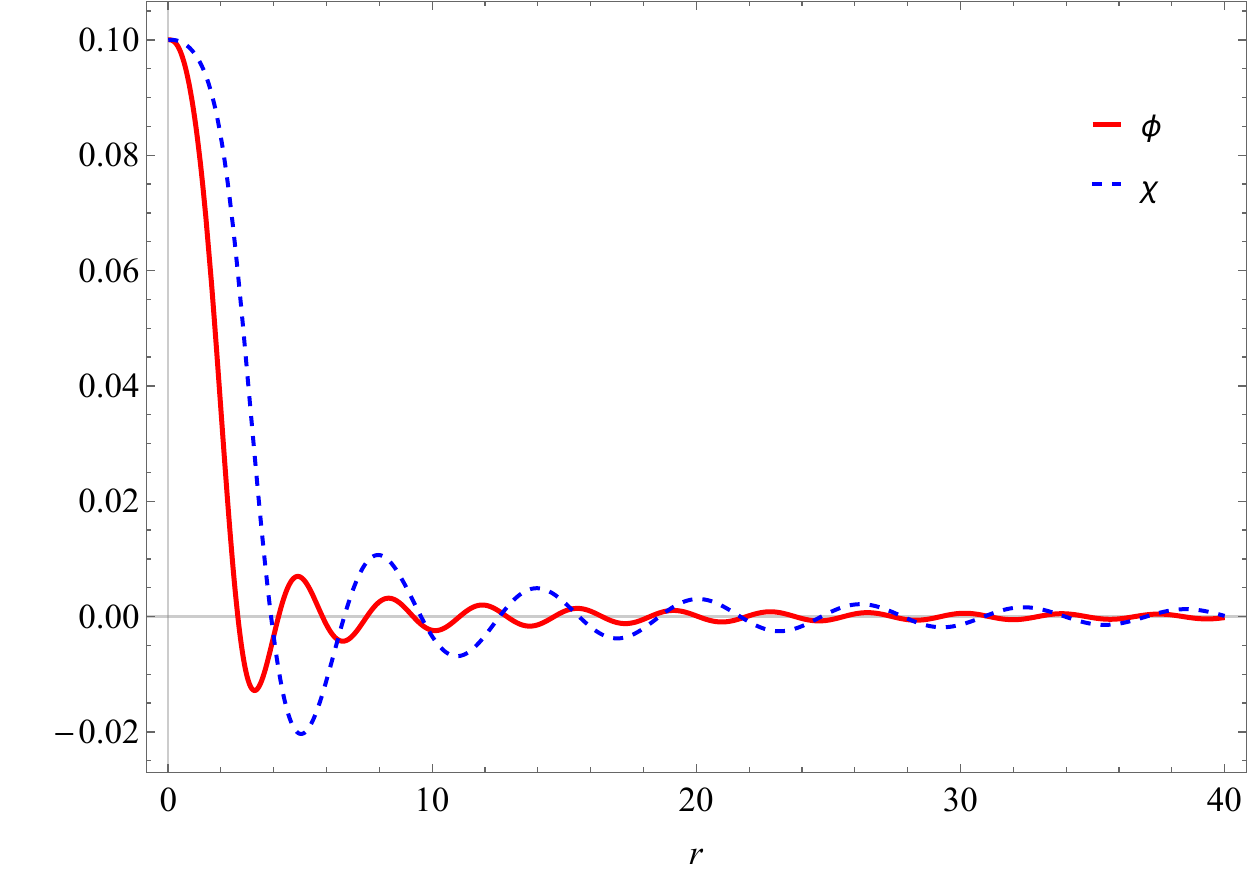}
}
\caption{These graphs are complementary of the Fig.~\ref{figs5} for the network around but inside the star.}
\label{figs-app-inside-2}
\end{figure}


\begin{figure}[!h]
\centering
\subfigure[{\bf \, Surface}-First case: The radial metric functions $C_0, C_2, C_4$. ]{
\includegraphics[width=.31\textwidth]{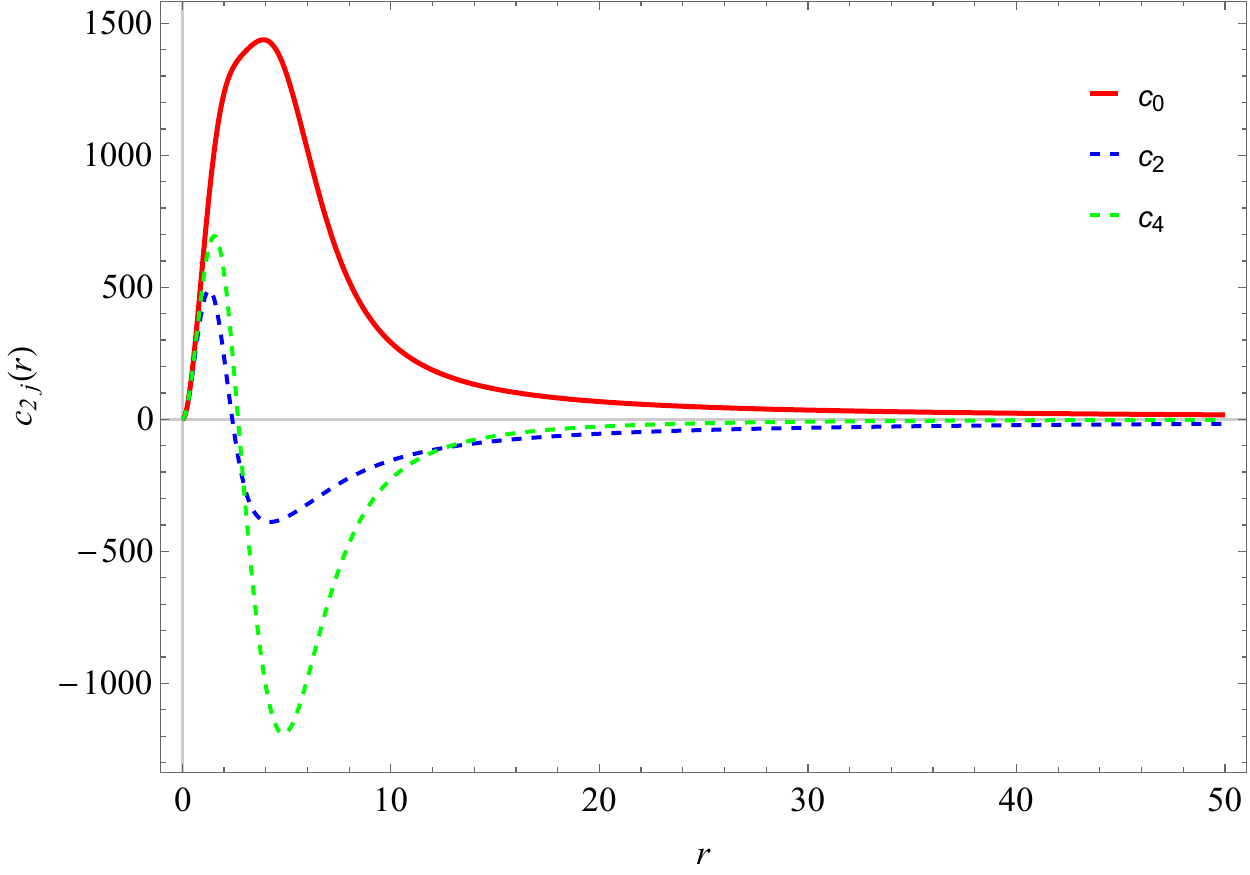}
}
\subfigure[{\bf \, Surface}-First case: The first two modes of the scalar field $\sigma$.]{
\includegraphics[width=.31\textwidth]{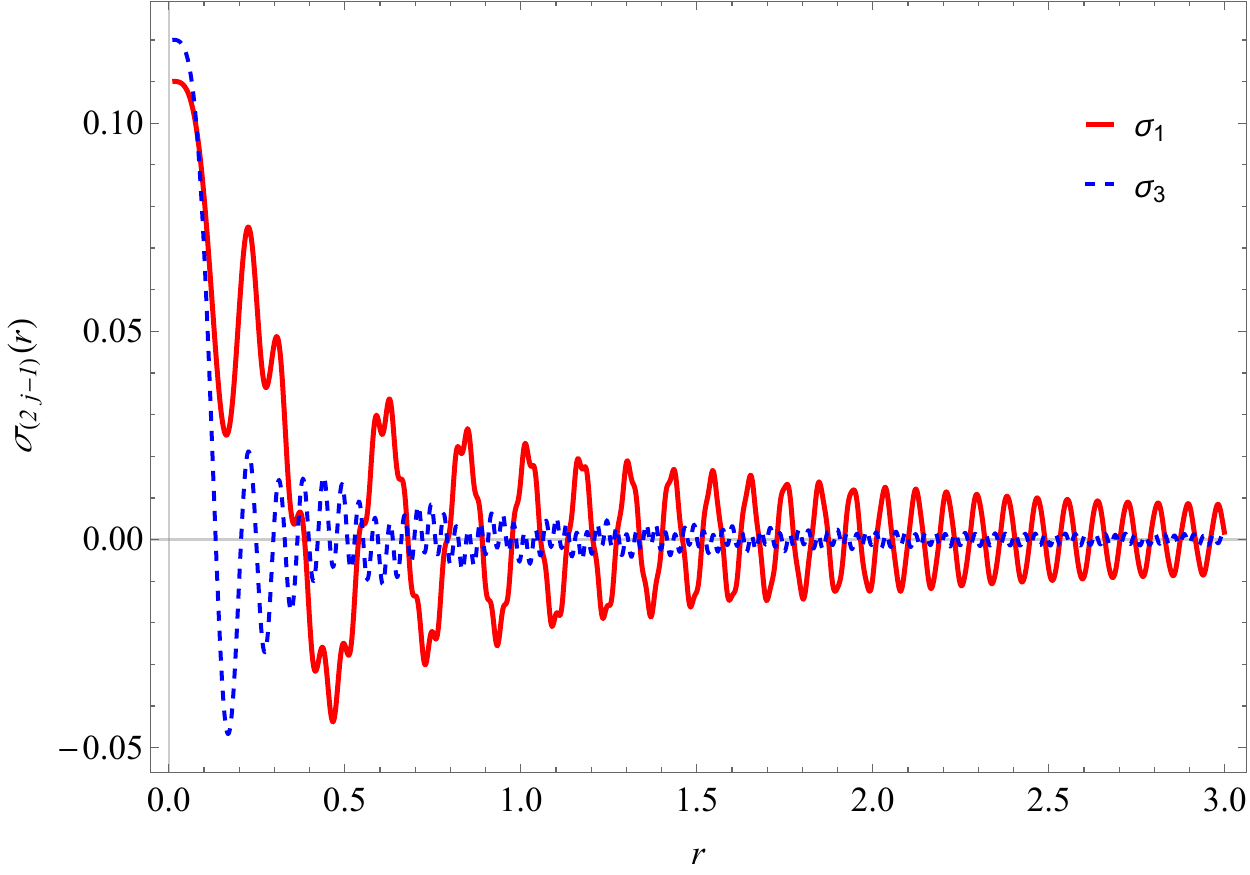}
}
\subfigure[{\bf \, Surface}-First case: The fields $\phi$ and $\chi$ inside the oscillaton.]{
\includegraphics[width=.31\textwidth]{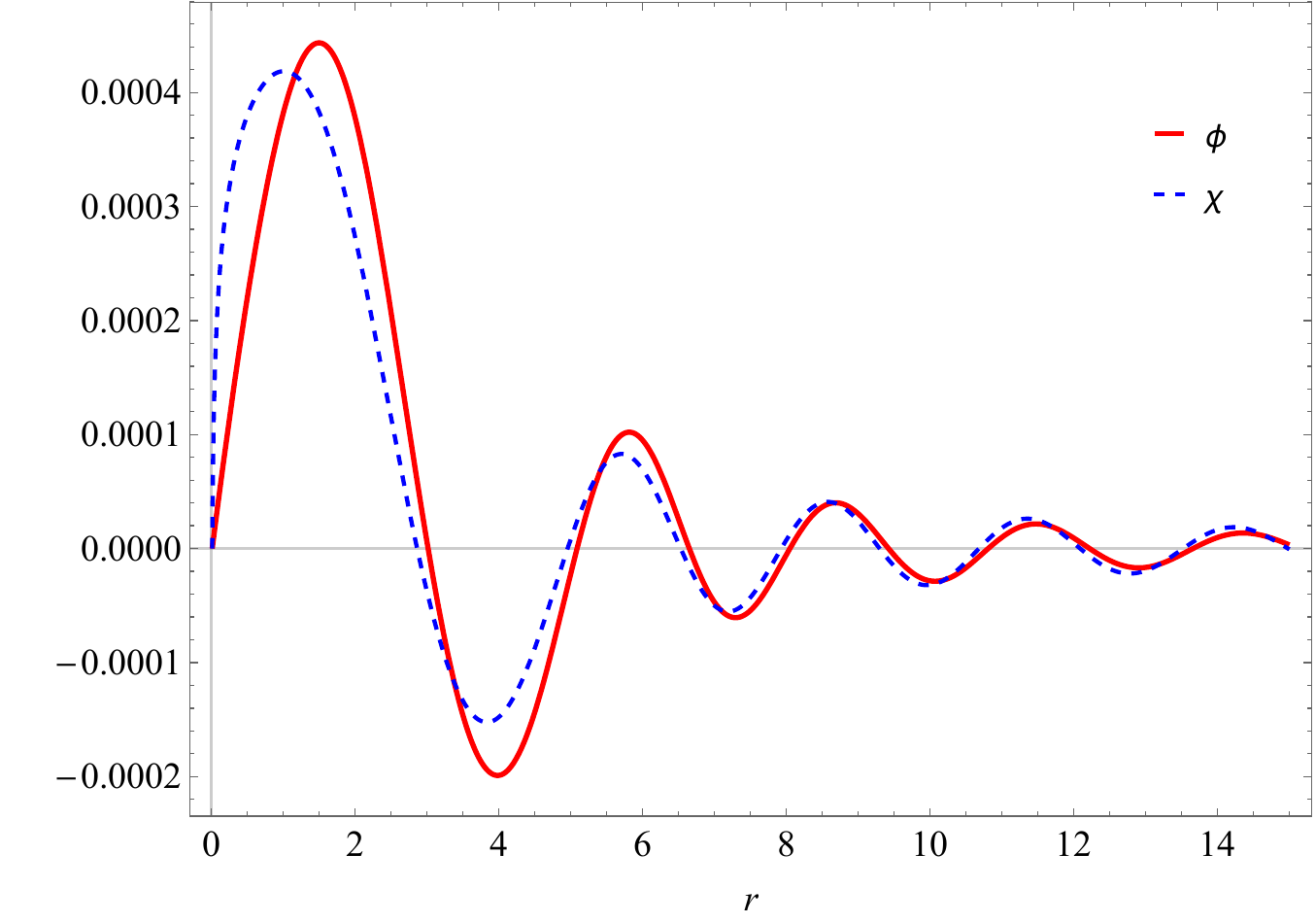}
}
\caption{These graphs are complementary of the Fig.~\ref{figs3} for the network on the surface of the star.}
\label{figs-app-surface-1}
\end{figure}

\begin{figure}[!h]
\centering
\subfigure[{\bf \, Surface}-Second case: The radial metric functions $C_0, C_2, C_4$. ]{
\includegraphics[width=.31\textwidth]{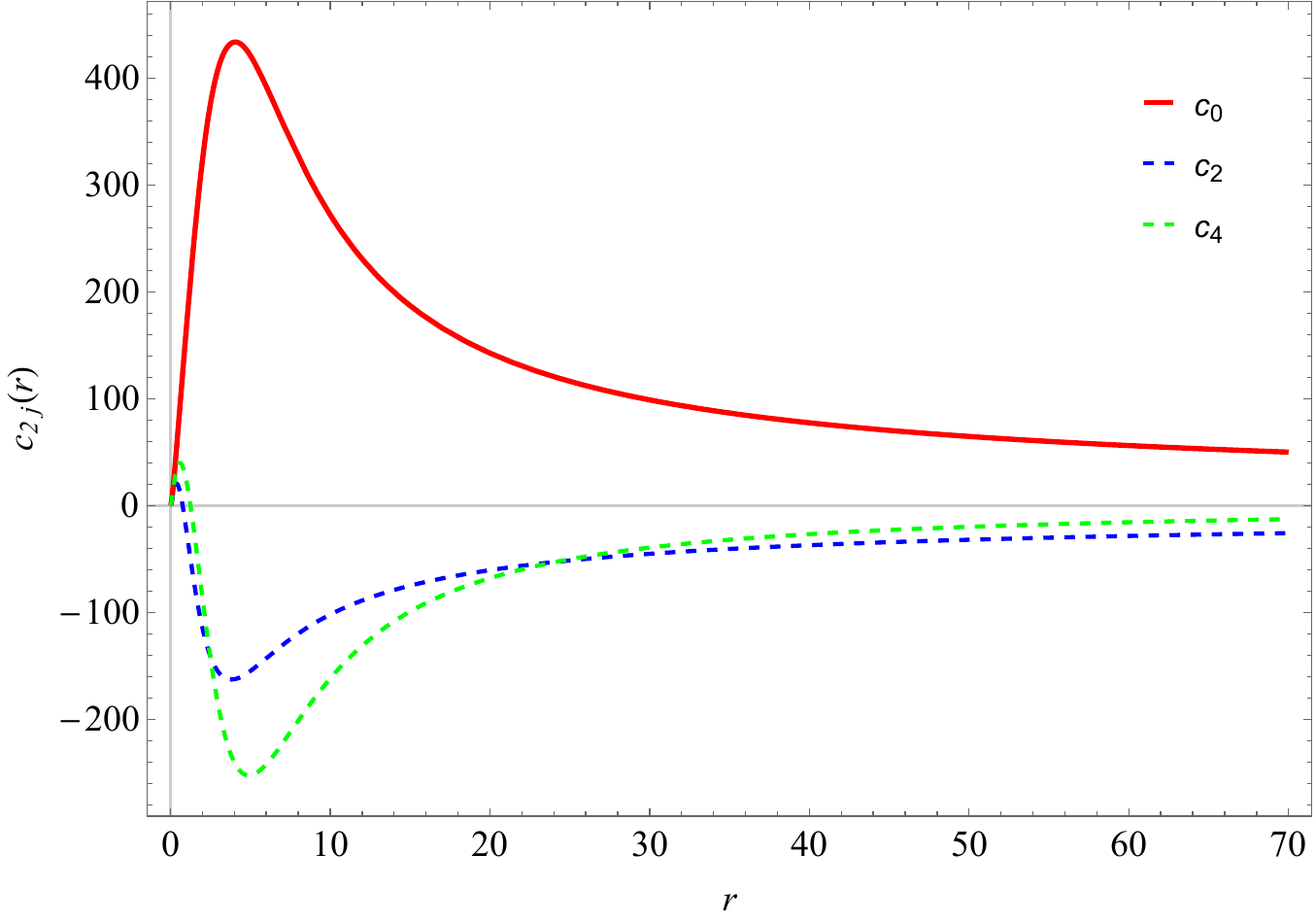}
}
\subfigure[{\bf \, Surface}-Second case: The first two modes of the scalar field $\sigma$.]{
\includegraphics[width=.31\textwidth]{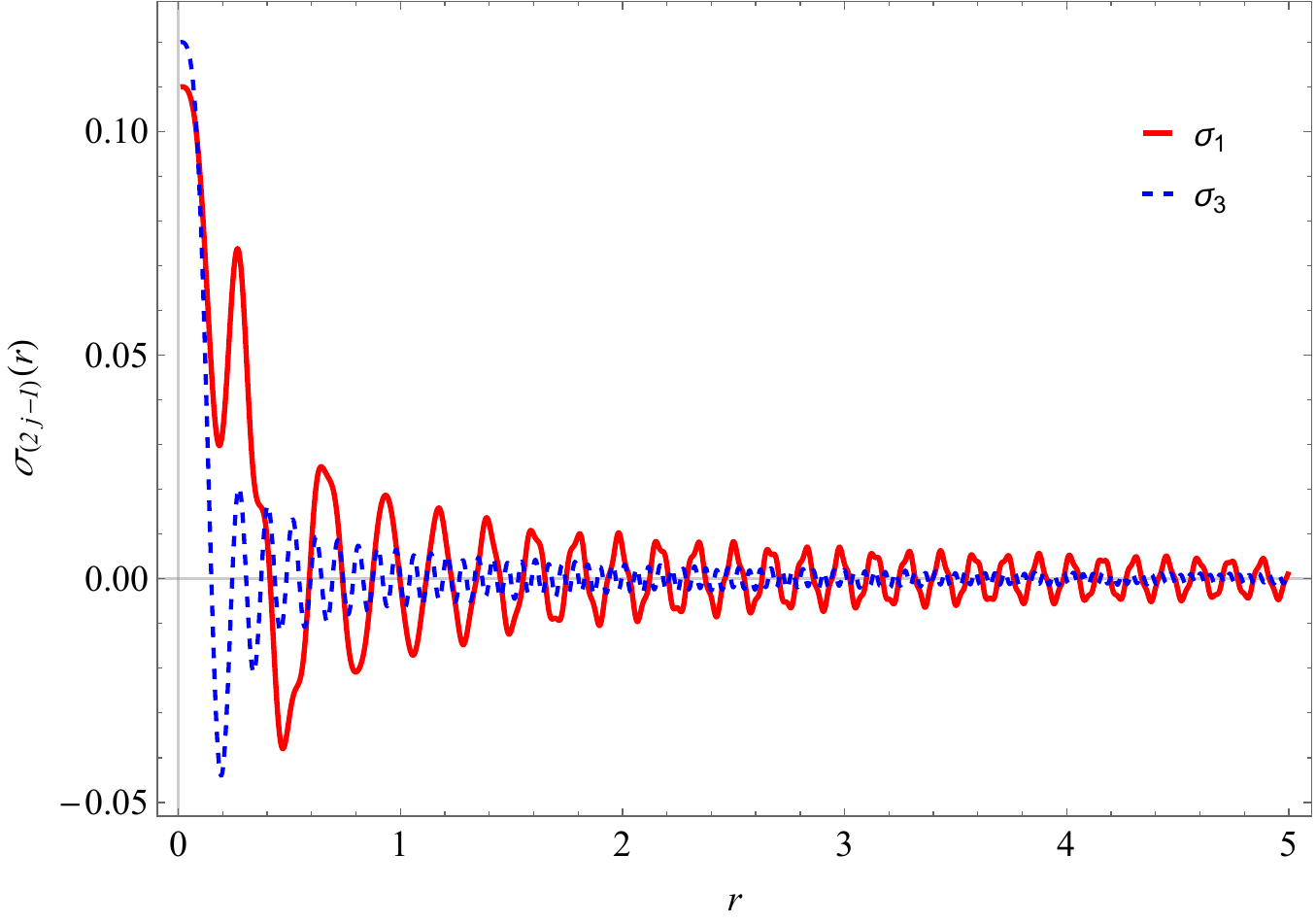}
}
\subfigure[{\bf \, Surface}-Second case: The fields $\phi$ and $\chi$ outside the oscillaton.]{
\includegraphics[width=.31\textwidth]{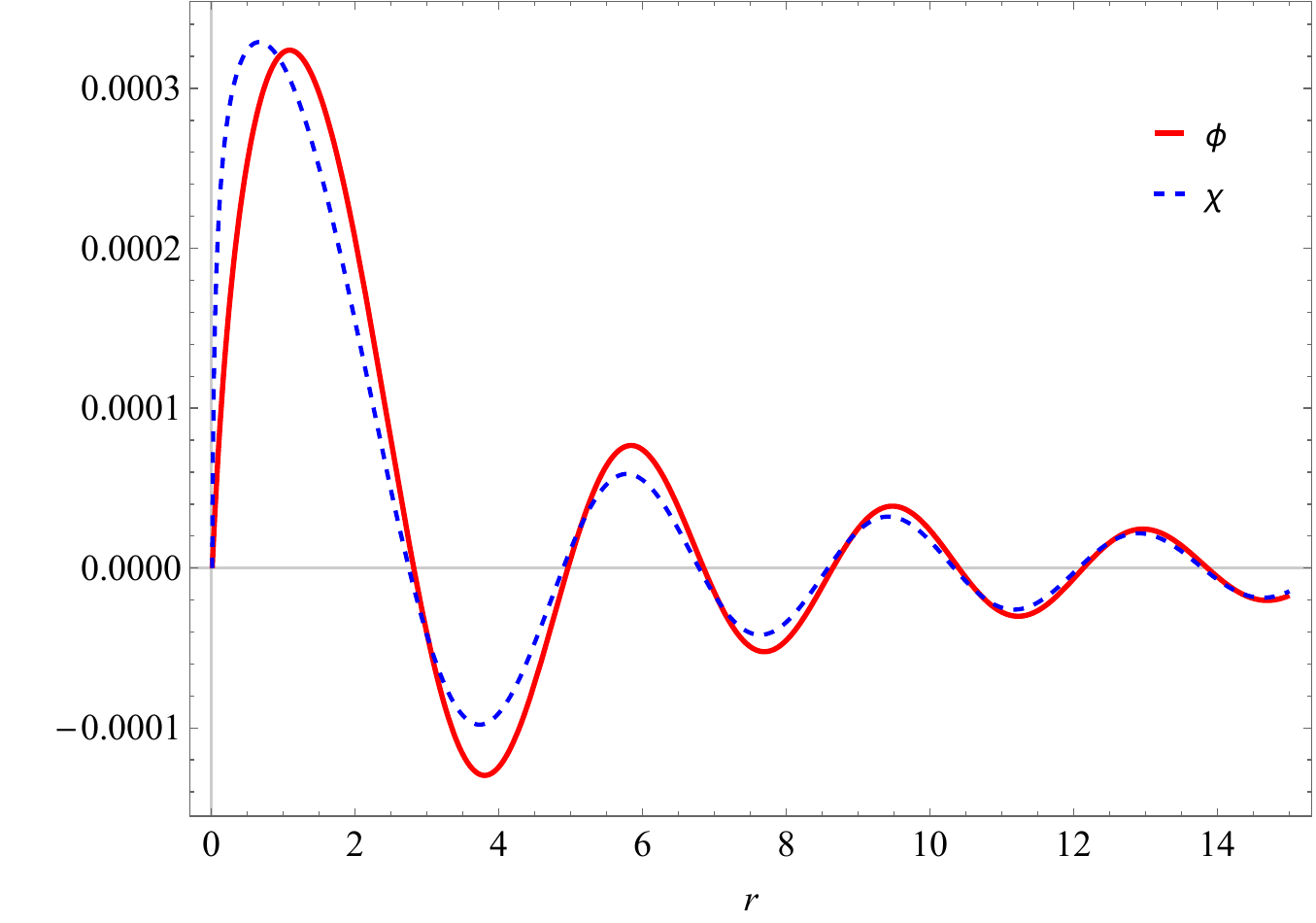}
}
\caption{These graphs are complementary of the Fig.~\ref{figs5} for the network on the surface of the star.}
\label{figs-app-suface-2}
\end{figure}


\begin{figure}[!h]
\centering
\subfigure[{\bf \, Outside}-First case: The radial metric functions $C_0, C_2, C_4$. ]{
\includegraphics[width=.31\textwidth]{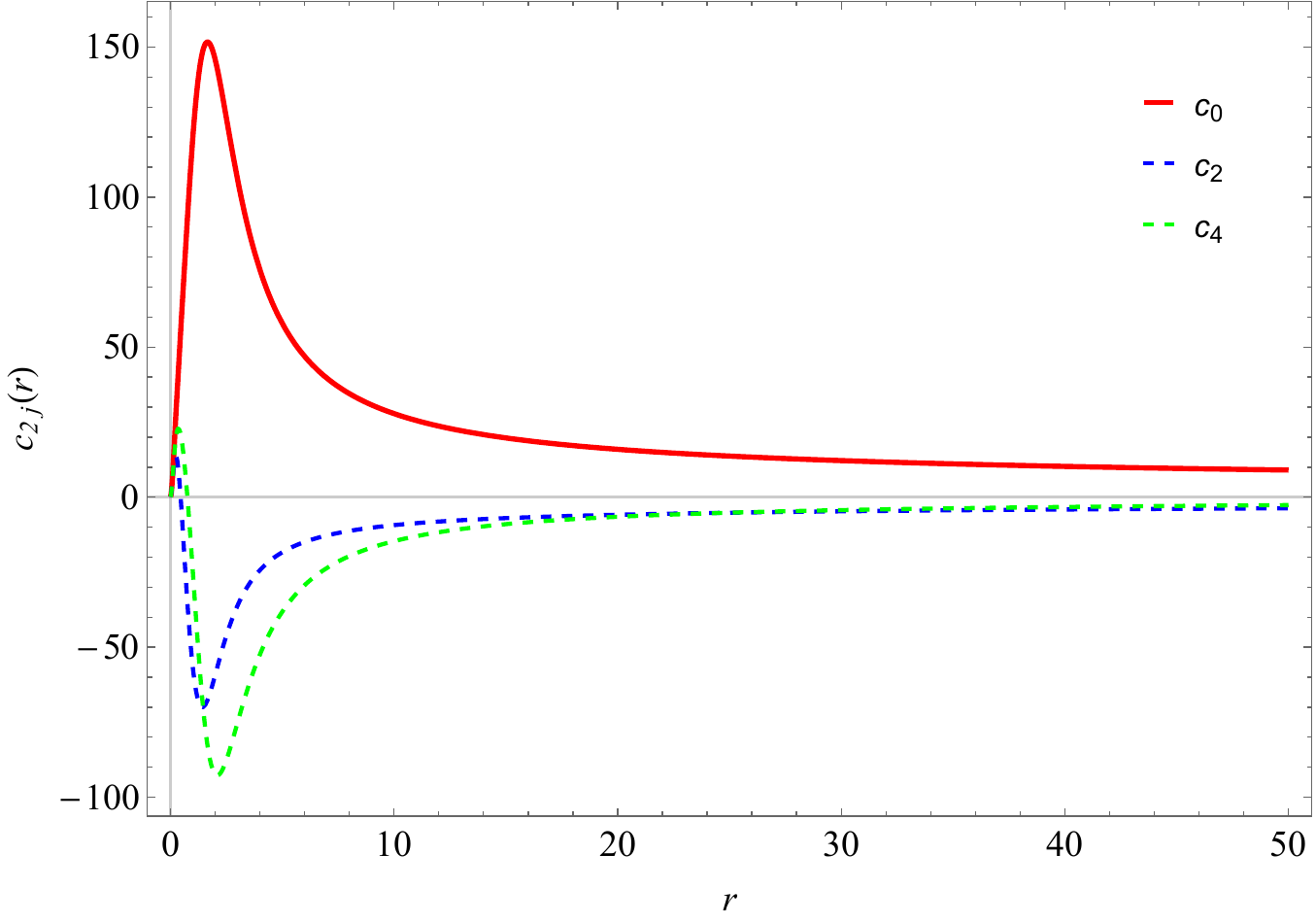}
}
\subfigure[{\bf \, Outside}-First case: The first two modes of the scalar field $\sigma$.]{
\includegraphics[width=.31\textwidth]{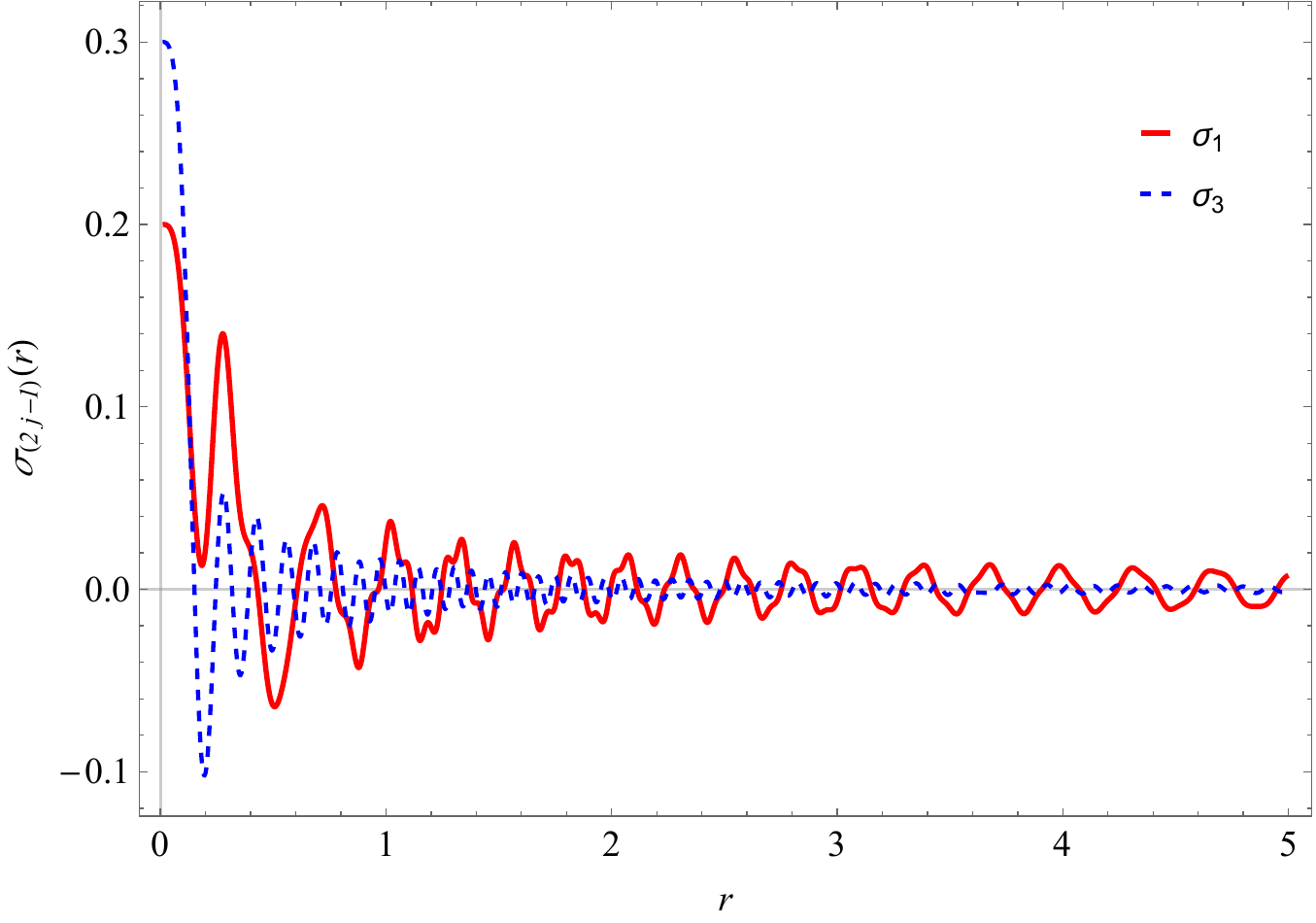}
}
\subfigure[{\bf \, Outside}-First case: The fields $\phi$ and $\chi$ outside the oscillaton.]{
\includegraphics[width=.31\textwidth]{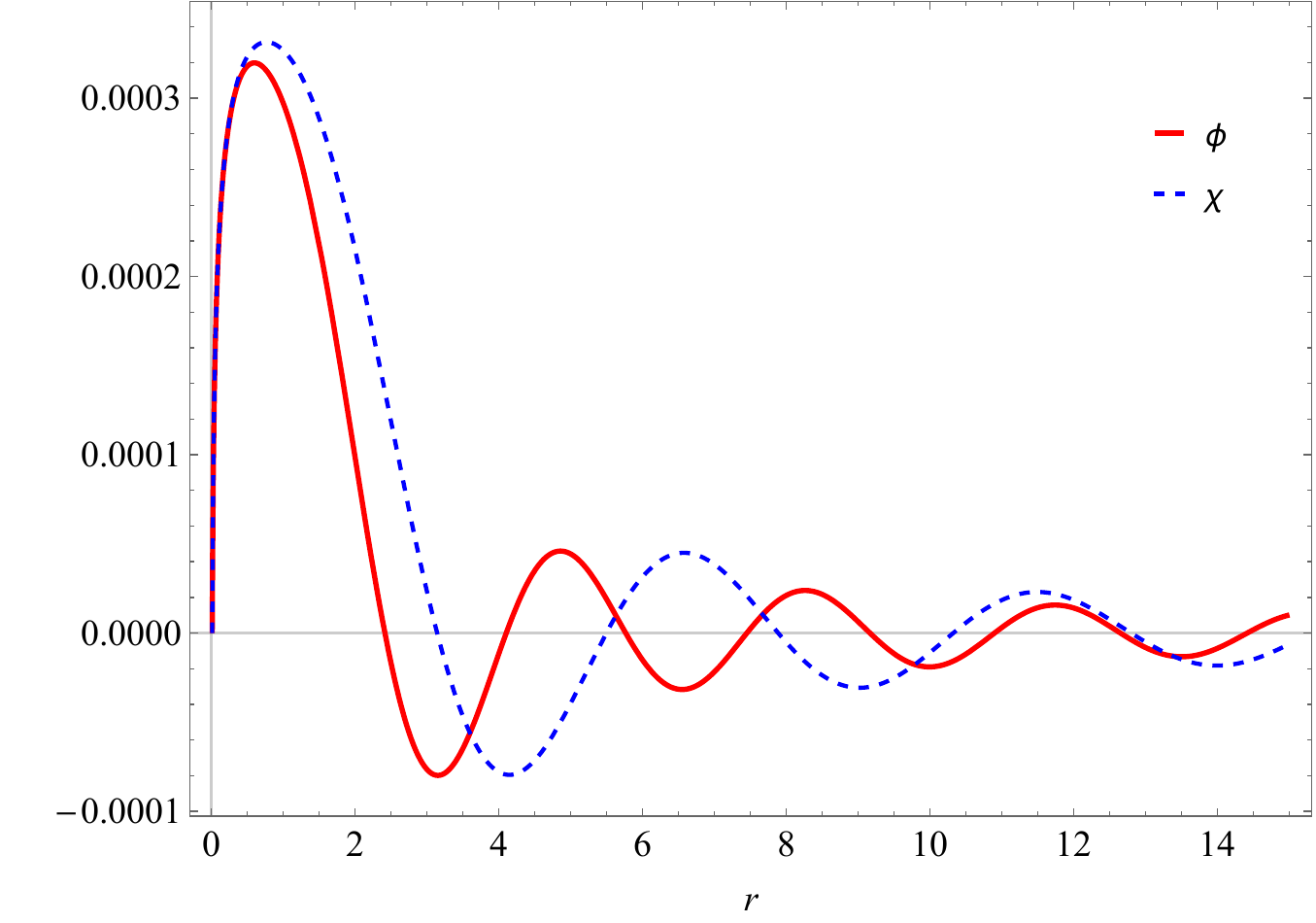}
}
\caption{These graphs are complementary of the Fig.~\ref{figs3} for the network around but outside the star.}
\label{figs-app-outside-1}
\end{figure}

\begin{figure}[!h]
\centering
\subfigure[{\bf \, Outside}-Second case: The radial metric functions $C_0, C_2, C_4$. ]{
\includegraphics[width=.31\textwidth]{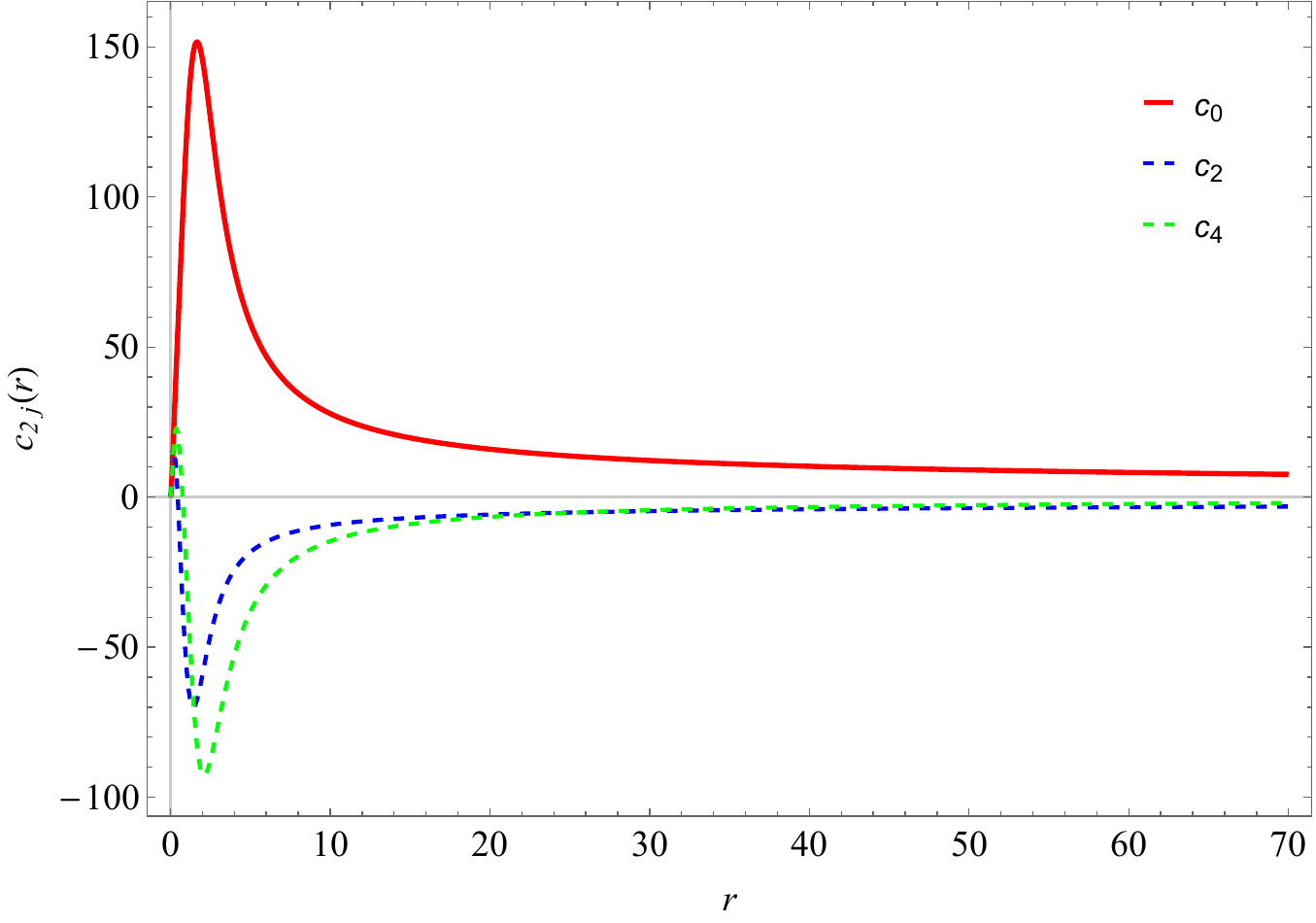}
}
\subfigure[{\bf \, Outside}-Second case: The first two modes of the scalar field $\sigma$.]{
\includegraphics[width=.31\textwidth]{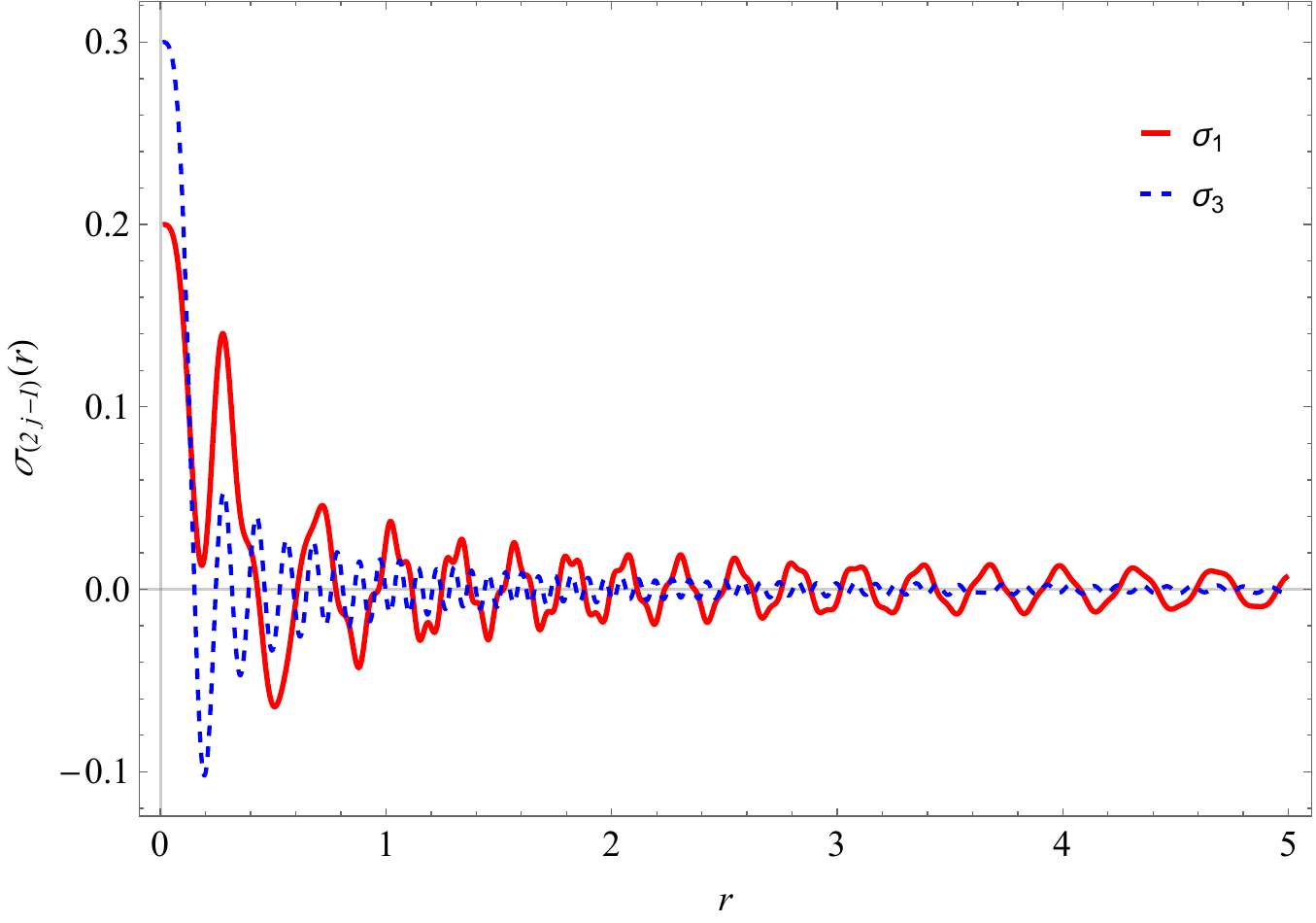}
}
\subfigure[{\bf \, Outside}-Second case: The fields $\phi$ and $\chi$ outside the oscillaton.]{
\includegraphics[width=.31\textwidth]{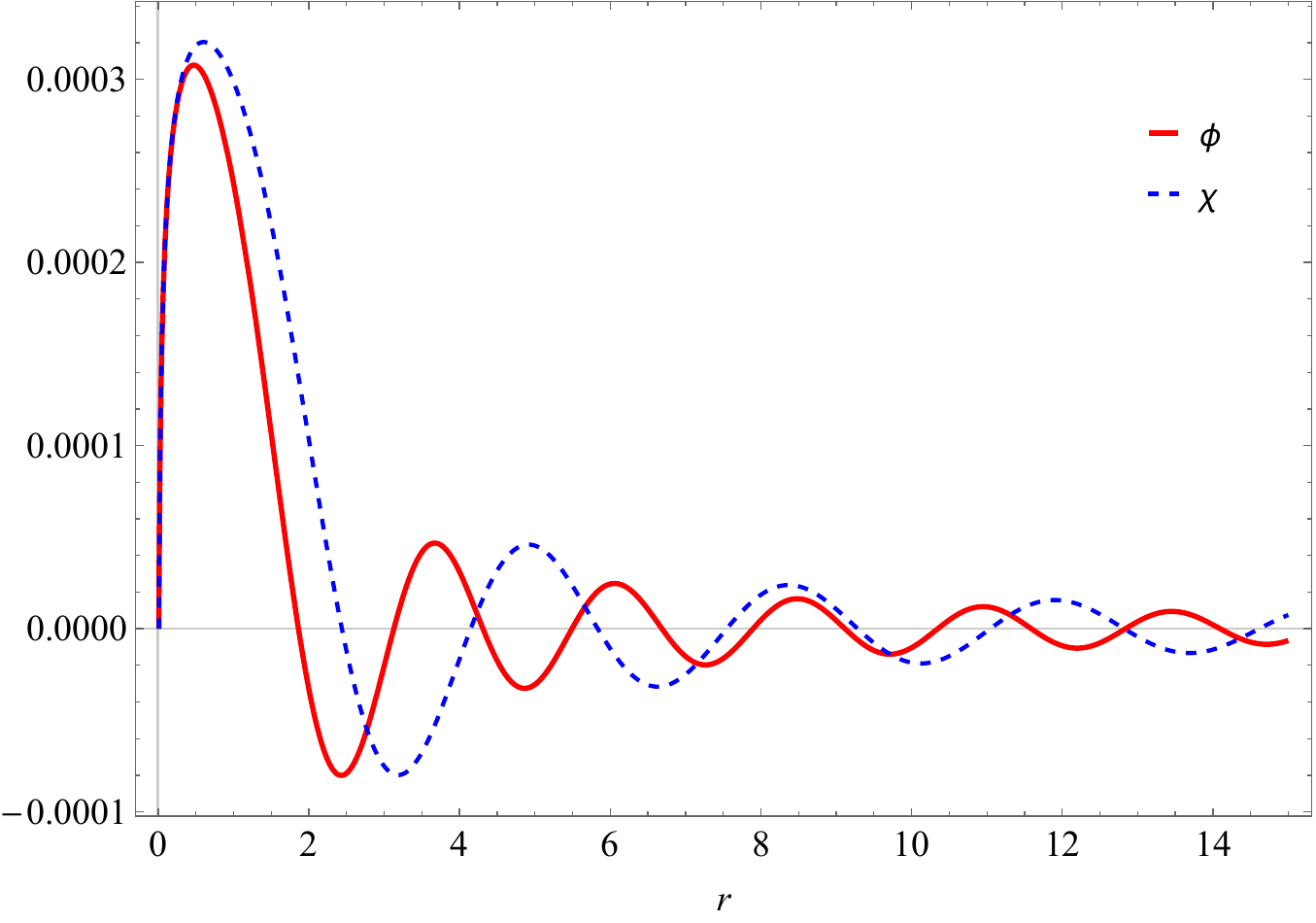}
}
\caption{These graphs are complementary of the Fig.~\ref{figs5} for the network around but outside the star.}
\label{figs-app-outside-2}
\end{figure}



\newpage
\begin{thebibliography}{99}
\bibitem{Lee}{T.D. Lee}.  \textit{Soliton stars and the critical masses of black holes.} Phys. Rev. D{\bf35}, 3637 (1987).
\bibitem{Lee and Pang}{T.D. Lee and Y. Pang}.  \textit{Fermion soliton stars and black holes.} Phys. Rev. D{\bf35}, 3678 (1987).
\bibitem{Friedberg}{R. Friedberg, T.D. Lee and Y. Pang}.  \textit{Scalar soliton stars and black holes} Phys. Rev. D{\bf35}, 3658 (1987).
\bibitem{C.W}{C.W. Misner  K.S. Thorne and J.A. Wheeler, \textit{Gravitation}} (Feeman, San Francisco, 1973).
\bibitem{R.M}	{R. M. Wald. \textit{General Relativity }} (University of Chicago Press, Chicago, 1984).
\bibitem{Friedberg2}{R. Friedberg, T.D. Lee and Y. Pang}.  \textit{Mini-Soliton stars.} Phys. Rev. D{\bf35}, 3640 (1987).
\bibitem{Francisco}{F.A. Brito and D. Bazeia}.  \textit{Network of domain walls on soliton stars.} Phys. Rev. D{\bf64}, 065022 (2001); hep-th/0105296.
\bibitem{sutcliffe} P. Sutcliffe. \textit{Domain wall network on solitons}. Phys. Rev. D{\bf68} (2003) 085004; [hep- th/0305198].
\bibitem{PBH} 
  M.~Y.~Khlopov.
  \textit{Primordial Black Holes.}
  Res.\ Astron.\ Astrophys.\  {\bf 10}, 495 (2010)
  doi:10.1088/1674-4527/10/6/001
  [arXiv:0801.0116 [astro-ph]].

\bibitem{PBH-QCD} 
  K.~Jedamzik.
   \textit{Could MACHOS be primordial black holes formed during the QCD epoch?}
  Phys.\ Rept.\  {\bf 307}, 155 (1998)
  doi:10.1016/S0370-1573(98)00067-2
  [astro-ph/9805147].
  
\bibitem{PBH-QCD2} 
  P.~Widerin and C.~Schmid.
   \textit{Primordial black holes from the QCD transition?}
  astro-ph/9808142.
  

\bibitem{T.D}{T.D. Lee}. \textit{Example of four-dimensional soliton solutions and abnormal nuclear states.} Phys. Rep. {\bf23} C, 254 (1976).
\bibitem{Coleman}{S. Coleman}. \textit{Q-balls.} Nucl. Phys. {\bf B 262}, 263 (1985).
\bibitem{E.S}{E. Seidal and W.-M. Suen}. \textit{Dynamical evolution of boson stars: Perturbing the ground state.} Phys. Rev. D{\bf42}, 384 (1990).
\bibitem{W.M}{E. Seidal and W.-M. Suen}. \textit{Oscillating soliton stars.} Phys. Rev. Lett.  {\bf72}, 1659 (1991).

\bibitem{D.F}{D. Bazeia and F. A. Brito}. \textit{Entrapment of network of domain walls.} Phys. Rev. D{\bf62}, 101701 (R) (2000); hep-th/0005045.

\bibitem{stephen} S. Owusu, {\it Estrelas de solitons oscilantes com rede de paredes de dom\'\i nios}, Campina Grande, 2017 (Master's Thesis in Portuguese): http://dspace.sti.ufcg.edu.br:8080/jspui/handle/riufcg/2354.

\bibitem{PBH-DM} 
  B.~Carr, F.~Kuhnel and M.~Sandstad.
   \textit{Primordial Black Holes as Dark Matter.}
  Phys.\ Rev.\ D {\bf 94}, no. 8, 083504 (2016)
  doi:10.1103/PhysRevD.94.083504
  [arXiv:1607.06077 [astro-ph.CO]].

\bibitem{LIGO}B.P. Abbott et al.\textit{Observation of Gravitational Waves from a Binary Black Hole Merger.} Phys. Rev. Lett. {\bf116}, 061102 (2016); arXiv:1602.03837 [gr-qc].

\bibitem{M. Al}{M. Alcubierre et al}. \textit{Numerical studies of $\phi^2$ oscillatons.} Class. Quant. Grav. {\bf20}, 2883 (2003); gr-qc/0301105.


\bibitem{living} S. Liebling and C. Palenzuela,  Living Reviews in Relativity, {\bf26}  (2023); DOI:10.1007/s41114-023-00043-4.


\bibitem{Balakrishna}{J. Balakrishna, R. Bondarescu, G. Daues and M. Bondarescu. \textit{Numerical Simulations of Oscillating Soliton Stars: Excited states in Spherical Symmetry and ground state evolutions in 3D}}. Phys. Rev. D{\bf77}, 024028 (2008); [arxiv: 0710.4131[gr-qc]].
\bibitem{L.A}{ L.A. Urena-Lopez, T. Matos, R. Beccerril}. \textit{Inside oscillatons.} Class. Quant. Grav. {\bf19}, 6259 (2002).

\bibitem{L.A.2} 
  L.A. Urena-Lopez, S.~Valdez-Alvarado and R.~Becerril.
  \textit{Evolution and stability $\phi^4$ oscillatons}.
  Class.\ Quant.\ Grav.\  {\bf 29}, 065021 (2012).
  doi:10.1088/0264-9381/29/6/065021.
  
\bibitem{T. Kodama}{ T. Kodama, L.C.S de Oliveira, and F.C. Santos}. \textit{Properties of a general-relativistic kink solution.} Phys. Rev. D{\bf19}, 3576 (1979).

 \bibitem{stability2}F.V. Kusmartsev, E.W. Mielke, and F.E. Schunck, {\it Gravitational stability of boson stars}, Phys. Rev. D {\bf43}, 3895 (1991).


\bibitem{ceres} 
  Y.~Brihaye, A.~Cisterna and C.~Erices.
   \textit{Boson stars in biscalar extensions of Horndeski gravity.}
  Phys.\ Rev.\ D {\bf 93}, no. 12, 124057 (2016)
  doi:10.1103/PhysRevD.93.124057
  [arXiv:1604.02121 [hep-th]].

\bibitem{Baibhav:2016fot} 
  V.~Baibhav and D.~Maity. 
  \textit{Boson Stars in Higher Derivative Gravity}.
  Phys.\ Rev.\ D{\bf 95}, no. 2, 024027 (2017)
  doi:10.1103/PhysRevD.95.024027
  [arXiv:1609.07225 [gr-qc]].

\bibitem{D.To}{D. Bazeia and F.A. Brito}. \textit{Bags, junctions, and network of BPS and non-BPS defects.} Phys. Rev. D{\bf61}, 105019 (2000).

\bibitem{Kolb}{E. W. Kolb and M.S Turner. \textit{The Early Universe. }}(Addison-Wesly, 1990).


\end{thebibliography}
\end{document}